\documentclass[11pt,a4paper]{scrartcl}
  \usepackage[pdf]{preamble}
  \usepackage{abstract}
  \title{\vspace{-2cm}Engines of Parsimony: Part I}
  \subtitle{Limits on Computational Rates in Physical Systems}

  \makeatletter
    \hearley@captionsetup{format=plain}
  \makeatother

  \usepackage{calc}
  \usetikzlibrary{arrows.meta}
  \makeatletter
  \newlength\hearley@arrlen
  \def\hearley@arr#1#2{\ensuremath{%
    \setlength\hearley@arrlen{\widthof{\ensuremath{#2}}}%
    \overset{\kern#10.1ex\raisebox{-0.3ex}[0pt][0pt]{%
      \clap{\tikz \draw[-{Stealth[length=0.4ex,width=0.4ex]}]%
        (0,0) -- (#1\hearley@arrlen,0);}}}{#2}}}
  \def\fwd{\hearley@arr{}}
  \def\bwd{\hearley@arr{-}}
  \def\smfwd#1{\smash{\fwd{#1}}}
  \def\smbwd#1{\smash{\bwd{#1}}}
  \makeatother
\endofdump

\begin{document}
\maketitle

\begin{figure}[h!]
  \vspace{-1.5cm}
  \centering
  \includegraphics[width=.9\textwidth]{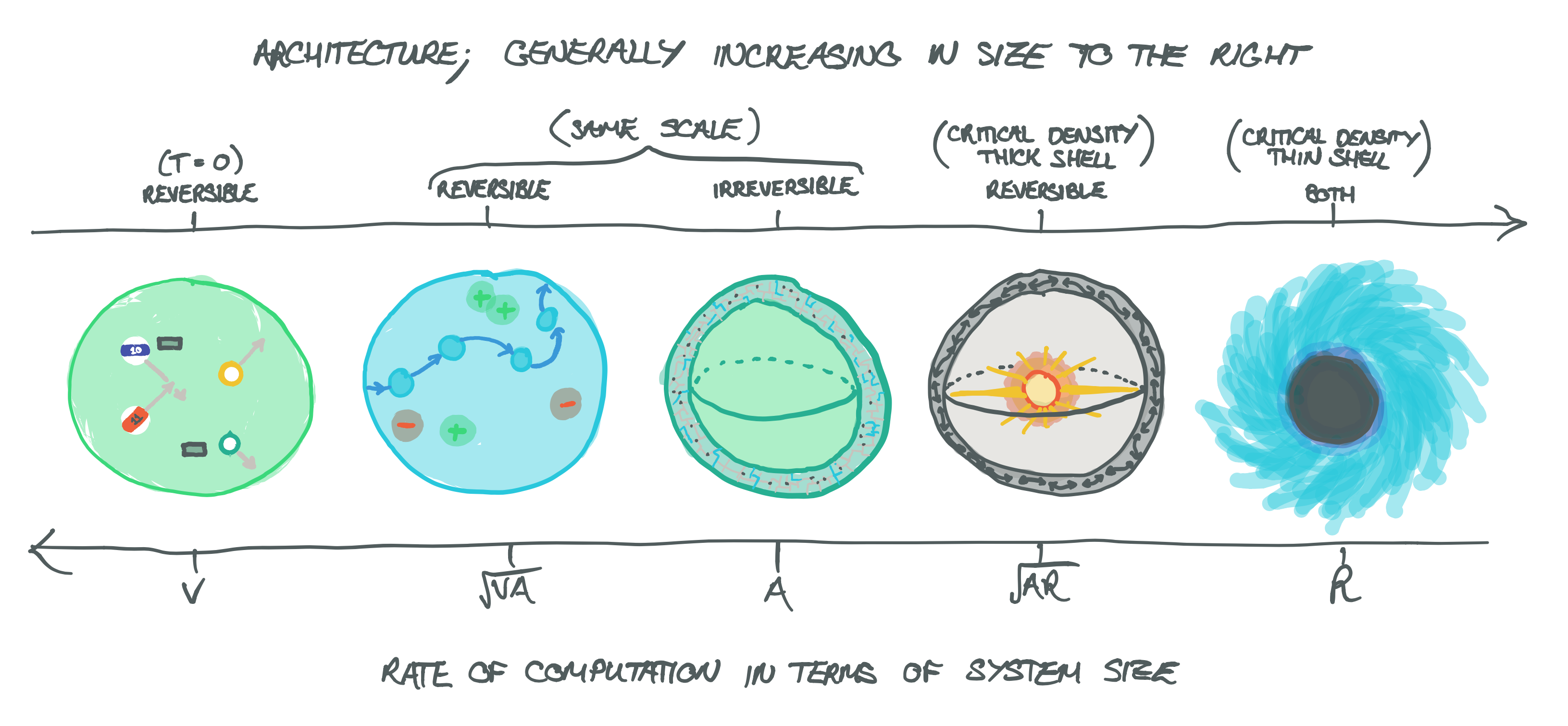}
  \captionsetup{singlelinecheck=off,type=figure}
  
    \newlength\figprefixlen\settowidth{\figprefixlen}{Figure~99:~}
    
  \caption[An illustration of the primary results of this paper]{%
    An illustration of the primary results of this paper. Consider a spherical region of computational matter with radius~$R$, convex surface area~$A$ and enclosed volume~$V$. The expressions indicate how the computational rate of each architecture scales, proportionally. From left to right, these regimes are:%
    \\[1em]%
    \begin{tabularx}{.99\textwidth-\figprefixlen}{rX}%
        $V:$ & reversible computers at absolute zero; \\%
        $\sqrt{AV}:$ & reversible computers at finite temperature; \\%
        $A:$ & irreversible/canonical computers; \\%
        $\sqrt{RA}:$ & critical density (thick shell on the cusp of gravitational collapse); \\%
        $R:$ & critical density (thin shell on the cusp of gravitational collapse).%
    \end{tabularx}%
  }
  \label{fig:scaling}
\end{figure}

\renewcommand{\abstractname}{Lay Summary}
\begin{abstract}
  In recent years, unconventional forms of computing ranging from molecular computers made out of DNA to quantum computers have started to be realised. Not only that, but they are becoming increasingly sophisticated and have a lot of potential to influence the future of computing. One interesting class of unconventional computers is that of reversible computers, which includes quantum computers. Reversible computing---wherein state transitions must be invertible, and therefore must conserve information---is largely neglected outside of quantum computing, but as we show in this paper this neglect is highly detrimental to computational performance.
  
  In particular, we consider the maximum sustained computational performance---in terms of operations per second---that can be extracted from a given region of space, under fairly general assumptions. Namely, we assume the known laws of physics, and that one will need to supply the system with energy over time in order to keep it going. We show, similarly to early work by \textcite{frank-thesis}, that for any realisable computer reversible computers are strictly better than irreversible computers at any size. We also derive universal scaling laws describing just how much better a reversible computer could be compared to an irreversible computer, proving that the adiabatic \emph{Time-Proportionally Reversible Architectures} (TPRAs) of Frank are the best possible, and suggest a path to achieving this bound with molecular computers.
  
  To illustrate these results, summarised in \Cref{fig:scaling}, suppose we wish to build the most powerful computer we can within some spherical region of space, of radius $R$. If the computer is irreversible, such as conventional silicon-based processors, then it is found that essentially only the surface of this sphere can be used for computation and the interior must be empty or inert. This clearly limits the rate of computation proportional to $4\pi R^2$. The reason for this restriction is thermodynamic, arising from constraints on both supply of power and rejection of heat. If $R$ is very big, such that the computer threatens to collapse into a black hole, then it is found that the computational rate can now only scale proportional to $R$.
  
  For a reversible computer the situation is substantially improved. In principle, a reversible computer can operate without producing an increase in entropy and without requiring any input of energy, and so the rate of computation could scale with the volume. Unfortunately in practice this is not possible, and a more careful consideration of the system's entropy is required. The rate of computation is found to scale proportional to  $\frac{4}{\sqrt{3}}\pi R^{5/2}$, geometrically between the area and volume. For large $R$, general relativistic effects become significant and this falls to the more restrictive bound $R^{3/2}$. For even larger $R$ this eventually falls to proportional to $R$, coinciding with the irreversible computer. The reason for this additional threshold between scaling laws is that the sub-relativistic reversible computers---where gravitational effects on spacetime are negligible---are not taking full computational advantage of their volume due to thermodynamic constraints; as the system gets larger beyond the collapse threshold (when the geometry transitions to a thick shell rather than a sphere) the thermodynamic constraints gradually relax, allowing a scaling that `temporarily' exceeds the limit value of $R$.
  
  Therefore we see that at almost all scales, reversible computers substantially outperform irreversible ones, whilst at extremely small---typically sub-nanometer---and large---order of the visible universe in size---scales they coincide.
\end{abstract}

\enlargethispage*{2\baselineskip}
\renewcommand{\abstractname}{Technical Abstract}
\begin{abstract}
  We analyse the maximum achievable rate of sustained computation for a given convex region of three dimensional space subject to geometric constraints on power delivery and heat dissipation. We find a universal upper bound across both quantum and classical systems, scaling as $\sqrt{AV}$ where $V$ is the region volume and $A$ its area, verifying and strengthening a result of \textcite{frank-thesis}. Attaining this bound requires the use of reversible computation, else it falls to scaling as $A$. By specialising our analysis to the case of Brownian classical systems, we also give a semi-constructive proof suggestive of an implementation attaining these bounds by means of molecular computers. For regions of astronomical size, general relativistic effects become significant and more restrictive bounds proportional to $\sqrt{AR}$ and $R$ are found to apply, where $R$ is its radius. It is also shown that inhomogeneity in computational structure is generally to be avoided. These results are depicted graphically in \Cref{fig:scaling}.
\end{abstract}

\clearpage

\section{Introduction}
\label{sec:intro}

The ubiquity of computer technology---and the increasing demands set upon it by intensive algorithms from fields such as machine learning, physics simulations, and cloud computing, among others---render the question of computer performance of considerable interest. Perhaps the most well known observation on computer performance was published by \textcite{moore-law}, who noticed the trend of microchip component density doubling every 18 months, later dubbed `Moore's law' in his honour. This law seemed to apply to many quantities in computing, including clock speed, \textit{FLOPS}\footnote{\textit{FLOPS}, or FLoating point Operations Per Second, is a common measure of supercomputer performance.} of the top 500 supercomputers combined, and inverse storage cost. A popular debate arose over when, if ever, Moore's law would stagnate; for clock speed, this point has come and gone, with consumer processor clock speeds frozen around \SIrange{3}{4}{\giga\hertz}. 

Whilst it is not surprising that contemporary technology may pose limits on computational performance, it is pertinent to ask whether the laws of physics impose hard bounds on performance. Indeed, our knowledge of quantum physics, thermodynamics and relativity reveals the answer to be affirmative. For a brief overview of some of these constraints see, for example, Lloyd's analysis~\cite{lloyd-ultimate}.

In this paper, we investigate how these bounds vary as a given computational system is scaled up. The results we find apply, with different constants of proportionality, for any given computational architecture. 
A computer constructed from a mix of architectures can be treated as a linear combination, and therefore these scaling results apply to any computer.
This builds upon the work of \textcite{frank-thesis}, proving that his notion of adiabatic \emph{Time-Proportionally Reversible Architectures} (TPRAs) are the best possible, confirming the relevant scaling results in the mesoscopic regime (as well as analysing scaling in very small and very large regimes in more detail), and excluding the possibility of architectures of lower dissipation. Moreover, the detailed calculations herein yield specific constants of proportionality for broad classes of reversible computational architectures.

There are many metrics by which one may wish to measure computational performance. In this paper, we shall investigate `speed' as defined by the rate of state transitions the computer is able to execute. As we are more concerned with the general form of scaling law rather than the specific constants for a specific architecture, the exact definition of `state' and `transition' are not too important. In general the most `obvious' definitions can be assumed, for example a conventional silicon-based computer would have as its states each possible value of all its registers, memory and any attached storage device, and as its transitions a single machine instruction\footnote{To be specific, we refer to the combination of a single op-code and its parameters.}. More rigorous definitions can be found in quantum mechanics, wherein a state would be an eigenstate of the computational basis Hamiltonian, and a transition would correspond to the complete evolution of the state between orthogonal eigenstates.

\begin{figure}[tb!]
  \centering
  \begin{subfigure}[b]{.45\linewidth}
    \includegraphics[width=.9\linewidth]{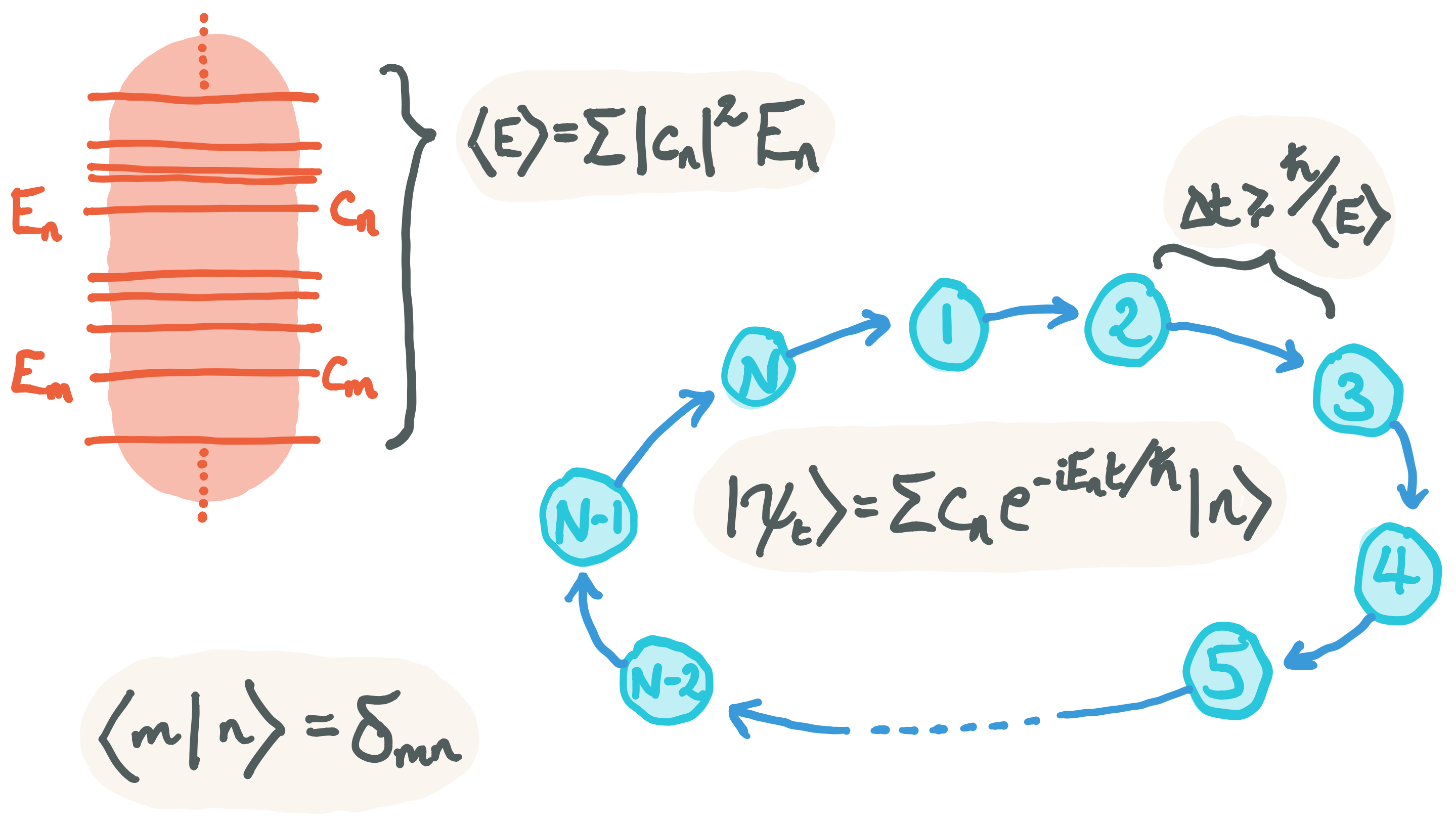}
    \vspace{1cm}
    \caption{}\label{fig:constraint-qm}
  \end{subfigure}%
  \begin{subfigure}[b]{.45\linewidth}
    \includegraphics[width=.9\linewidth]{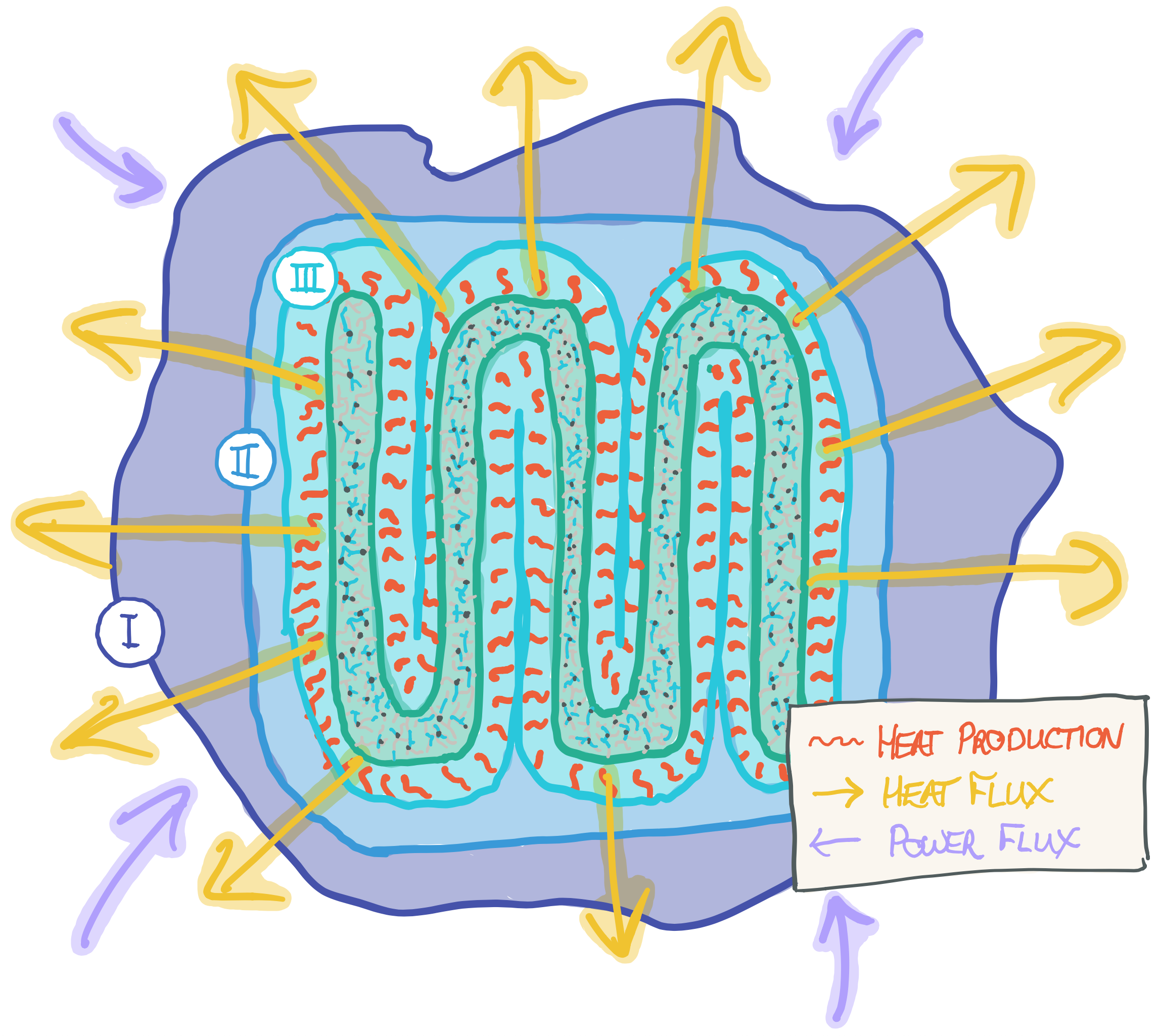}
    \caption{}\label{fig:constraint-td}
  \end{subfigure}
  \caption{An illustration of the three dominant physical constraints that apply to computational systems. Constraints (a) and (b) are described in \Cref{sec:intro} and (c) in \Cref{sec:gr}. (a) The quantum mechanical constraint bounding the minimum time for a state transition for a system of given average mass-energy $\evqty{E}$ where $E_0=0$. This illustration includes example energy levels of the eigenstates and a cyclic sequence of orthogonal states $\{1,\ldots,N\}$ which are each a superposition of the eigenstates, with the same average energy $\evqty{E}$. (b) A combination geometric-thermodynamic constraint. Thermodynamically, each computational operation generates entropy (generally heat) and this is bounded from below in the case of irreversible computation, but still non-zero in the reversible case. Geometrically, we can only dissipate this entropy at a rate scaling with the convex bounding surface, (II). The surface (III), whilst larger than (II), is not useful as the entropy flux must still pass through surface (II). Surface (I) is also larger, but the flux must first pass through surface (II). The green region is the computational system proper. At steady state there is an amortised balance between the entropy dissipation flux and input power flux. (continued)}\label{fig:constraint}
\end{figure}
\begin{figure}[tb!]\ContinuedFloat
  \centering
  \begin{subfigure}[b]{.9\linewidth}
    \includegraphics[width=.9\linewidth]{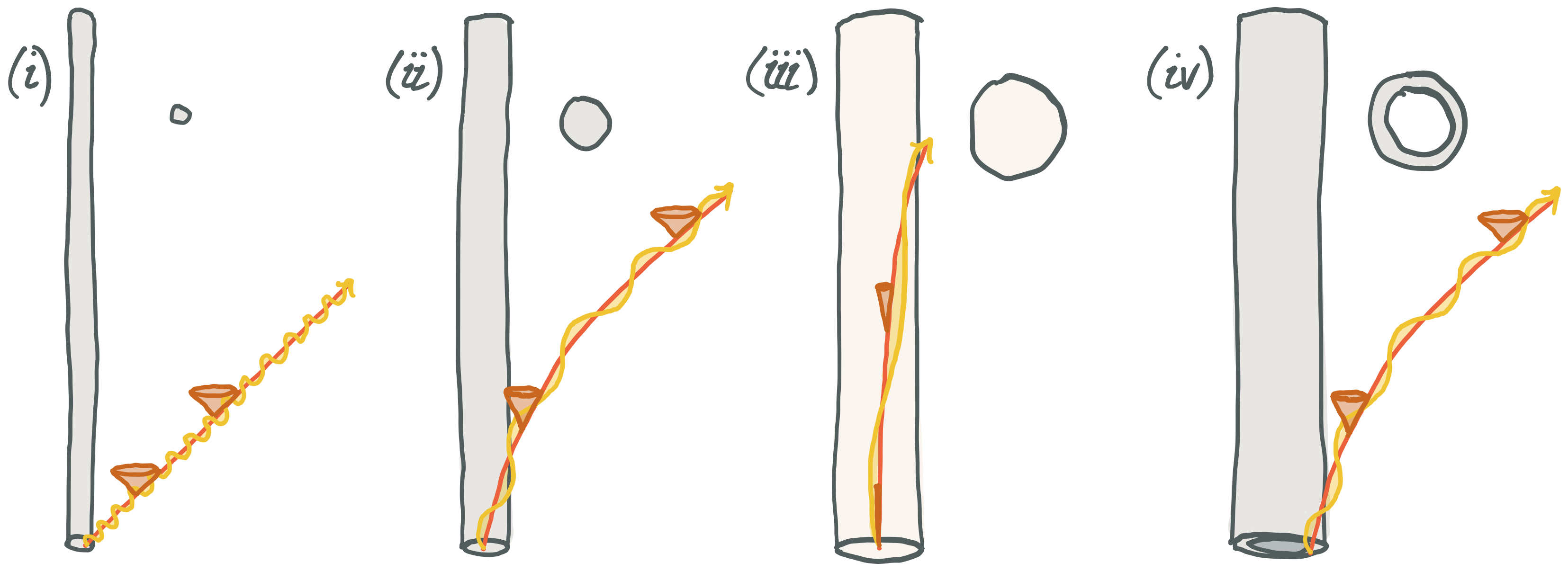}
    \caption{}\label{fig:constraint-gr}
  \end{subfigure}
  \caption{(continued) (c) The (general) relativistic constraint which informs the optimal mass distribution of the system. Each example computational system is shown as a space time diagram, with the horizontal axis corresponding to space and the vertical to time. Cones are light-cones showing the local causal structure and indicating spacetime curvature due to the stress-energy tensor. The emanating light waves correspond to the output of the computer. For small computers such as (i), general relativistic effects are negligible. For larger computers such as (ii), space is noticeably curved within the vicinity of the system and so the frequency of emitted photons is red-shifted, meaning that the computer appears slower to a distant observer. For computers on the cusp of gravitational collapse such as (iii)---even if the density is lowered to keep the Schwarzschild radius small, as indicated by the lighter shade of the system---light takes a very long time to escape, diverging to infinity as the system radius approaches $\frac98$ of the Schwarzschild radius. By keeping the total mass and radius the same but reconfiguring it into a spherical shell of the same density as (i,ii), as depicted in (iv), we can minimise the effect of time dilation to at most a threefold slowdown, and thus retain our qualitative power-law scaling.}
\end{figure}

Other performance metrics of interest might concern synchronisation between distinct computational elements within a parallel system, and interaction with some arbitrary non-equilibrium system such as a supply of additional memory resources. These shall be covered in parts II and III of this paper series, respectively~\cite{earley-parsimony-ii,earley-parsimony-iii}.

\para{Quantum Constraint on Computer Performance}

As a first approach to bounding computational performance, we turn to quantum mechanics. \textcite{bremermann-limit} gave a back of the envelope calculation in 1962, subsequently refined and elaborated upon by \textcite{margolus-levitin} and summarised in \Cref{fig:constraint-qm}, to show that a system with energy $E$ can change state at a maximum rate of $\nu\le E/h_P$ where $h_P$ is Planck's constant. Assuming this energy is due primarily to rest mass, this gives $\nu\le\SI{1.36e50}{\hertz\per\kilogram}$. Restricting our attention to a specific architecture as defined earlier, we assume a fixed density---or at least that the density is bounded from above---such that $\nu\le V\rho c^2/h_P$. This immediately implies that the maximum rate of computation scales with volume. 

\para{Thermodynamic and Geometric Constraint on Computer Performance}

This quantum limit assumes a closed computational system, requiring no external power source. As shall become clear, sustained processive computation without a driving force is for all practical purposes impossible. For conventional computing architectures, such as those inside your computer and smartphone, it is certainly impossible: this is because conventional computing is \textit{irreversible} in the sense that information is not generally conserved and so it is not possible to step backwards through a computation. This is in stark contrast to the laws of physics which are reversible and information conserving\footnote{This includes quantum mechanics; the Schrödinger equation specifies that the time evolution operator is unitary and hence invertible. Furthermore, it seems likely that the irreversible process of \textit{wavefunction collapse} is in fact a reversible one of progressive decoherence, as espoused by the Many-Worlds Interpretation~\cite{mwi}.}, and so in fact such models of computation are incompatible with the laws of physics. Rather, irreversible computation can only ever be realised in an approximate sense, leveraging thermodynamic principles to reject entropy to its environment. Irreversible computation is exhibited by, for example, non-invertible logic gates such as \texttt{AND}, variable mutation in which the previous state is lost, and merging of control flow. Even if it is possible in principle to reconstruct the previous computational state by some convoluted mechanism, this is insufficient as reversibility requires that a process can be readily rewound step-by-step. 

The resolution of this physical incompatibility relates to the connection between thermodynamics and information theory, formalised by \textcites{szilard-engine,landauer-limit} in the twentieth century. For every bit of information discarded, a commensurate entropy increase of at least $k\log 2$ is produced elsewhere, where $k$ is Boltzmann's constant\footnote{When this entropy takes the form of heat, each bit will generate on the order of $\sim\SI{e-21}{\joule}\sim\SI{e-2}{\electronvolt}$ at room temperature. Conventional computers typically expend a factor of around \num{e8} times this much for reasons relating to reliability and speed.}. More generally, whenever a quantity of information $I\,\text{bits}$ is discarded, an entropy increase of $\Delta S\ge kI\log 2$ manifests, with equality only in the limit of thermodynamic reversibility in which the discarding process takes an infinite period of time. If this entropy were allowed to remain in the computational system, it would eventually accumulate to such an extent that it would interfere with the well ordered operation of the computational mechanism. As a visceral example, should the entropy take the form of heat, $\Delta Q=T\Delta S$ where $T$ is the system temperature, the system will eventually transmute into a form incompatible with its function---such as melting or turning into a plasma---as the temperature becomes too great. Even if the system is physically heat-resistant, the increased entropy will result in errors so frequent that almost all the computational capacity is directed to error correction, slowing productive computation to a crawl.

In order to sustain computation then, it is necessary to remove the additional entropy at the same amortised rate as its production. In the case of heat, this is achieved by cooling the system; this process can be generalised to other forms of entropy, such as disorder in a spin bath~\cite{landauer-spin}, but we shall restrict out attention to heat for conceptual convenience. We must thus enact a flow of work into the system and heat out in order to sustain computation, but here we encounter a geometric constraint. This energy flow applies at every bounding surface, assuming that the heat is ultimately rejected to infinity. Consider a convex bounding surface of area $A$, and assume our technology of choice is capable of transporting energy up to a certain flux $\phi$, then the maximal power that can be exchanged with the system is given by $P\le\phi A$. As has been established, however, the rate of heat generation scales with the rate of computational transitions for an irreversible computer, and therefore the rate of computation is ultimately bounded by the system's convex surface area, rather than its volume as contended by the quantum mechanical limit. See \Cref{fig:constraint-td} for an illustration.

This is far more restrictive than the volumetric upper bound derived earlier, and suggests that for large computers only a subvolume equivalent to the outermost shell of the volume can perform useful computation with the remaining bulk relegated to dormant inactivity\footnote{The remaining bulk need not be entirely devoid of purpose. It may, for example, be used for storage and memory. It must however have negligible entropy generation, such as that resulting from structural deterioration due to cosmic rays, spontaneous tautomerisation, etc.}. That is, we take the more restrictive of the two bounds, which for small systems is the volumetric bound and for larger (irreversible) systems is this areametric bound.

\para{Ballistic Computation}

Does this areametric thermodynamic bound render the volumetric quantum bound inaccessible? A trivial exception is found at small scales, where the surface area to volume ratio becomes so large that the volumetric bound is in fact more restrictive than the areametric bound, and thus takes priority. To more robustly improve on the rate of computation, however, we must reduce the entropic cost associated with a computational transition. This necessitates conserving information during computation. Landauer, upon coming to this conclusion, argued that performing any useful computation under this limitation would be untenable as, instead of raising the entropy of the environment, one would fill the computer's memory with the discarded information instead. Fortunately, this objection was unfounded and a decade later \textcite{bennett-tm}%
, oft considered the Father of Reversible Computing,
explicated a viable model of reversible computation in his Reversible Turing Machine (RTM).
The key property of RTMs is that they require both the domains and codomains of their rules to be non-overlapping, rather than just their domains as is the case with standard Turing Machines\footnote{Turing Machines~\cite{turing-machine} were one of the first general models for describing an arbitrary computing machine, the other being the Lambda Calculus~\cite{lambda-calculus}. For an introduction to both, see textbooks on Automata Theory and Computability Theory.}.
Furthermore, he showed how any irreversible program could be efficiently simulated on such a reversible computer without excessive memory overhead, thus proving that reversible computing was equipotent with conventional computing. For a brief exposition of reversible programming, see \Cref{app:rev-comp}.

An idealised reversible computer would produce no entropy in its transitions. \textcite{fredkin-conlog} described such an idealised system, a physical model of reversible computing that could operate without being actively powered. This model consisted of a frictionless table upon which hard elastic billiard balls were projected. The balls would bounce off of each other and some strategically placed walls, with the precise configuration describing the resultant computation. As a testament to its capability, Fredkin's student, \textcite{ressler}, proposed the design of a fully programmable billiard ball computer, complete with arithmetic logic unit. Given that such a computer would be powered solely by its initial kinetic energy, the volumetric quantum bound would be attainable. Unfortunately, \textcite{bennett-rev} calculated that such a design would ultimately be infeasible as even the most distant and subtle influences would be sufficient to rapidly and completely thermalise the motions of the system:
\def\signed #1{{\leavevmode\unskip\nobreak\hfil\penalty50\hskip2em
  \hbox{}\nobreak\hfil{#1}%
  \parfillskip=0pt \finalhyphendemerits=0 \endgraf}}
\begin{quote}
  ``Even if classical balls could be shot with perfect accuracy into a perfect apparatus, fluctuating tidal forces from turbulence in the atmospheres of nearby stars would be enough to randomise their motion within a few hundred collisions. Needless to say, the trajectory would be spoiled much sooner if stronger nearby noise sources (e.g., thermal radiation and conduction) were not eliminated.''
  \signed{\textrm{ --- \textcite{bennett-rev}}}
\end{quote}
Such influences could in principle be suppressed by error correction mechanisms, but error correction corresponds to the discarding of excess entropy, and would seem to reconstitute the very issues we were trying to avoid. 

Our last refuge against unwarranted entropic influences is in quantum ground state condensates. Phenomena such as superfluidity and superconductivity arise in a sufficiently cooled system, wherein a macroscopic fraction of particles are found to inhabit their ground state, manifesting macroscopic quantum behaviours and a subsystem with vanishing entropy and temperature. The utility of such systems is manifold, but unfortunately it is doubtful that they can be exploited for dissipationless computation; the reason is that, in order to enact transitions to orthogonal quantum states, the Schrödinger equation requires us to prepare a superposition of (distinct) energy states which then rotates under the action of the Hamiltonian. The presence of non-ground eigenstates will necessarily result in a renewed vulnerability to thermodynamic perturbations. Nevertheless, such cold temperatures are not altogether useless, as unwanted fluctuations would still be significantly suppressed.

\para{Brownian Dynamics}

The fluctuations and dissipation that result from these thermodynamic perturbations lead to diffusion of the distribution in phase space, and hence loss of absolute control over it. Consequently there will be some level of unpredictability to the state of the system and its (generalised) velocities. In the limit of negligible control, the dynamics are (mostly) dominated by the noise source; this regime is referred to variously as Brownian or Langevin dynamics. In \Cref{sec:crn} it shall be shown how to leverage the minimal extant control over the system to make net progress, and to even robustly surpass irreversible computers in performance.

\para{Summary}

Whilst fully dissipationless computation is all but impossible, it is still possible in principle to tune the system so as to bring the entropic transition cost as close to zero as desired. In exchange, the rate of computation may itself be diminished. In the paper that follows, this compromise is evaluated across the range of known physics---from quantum to classical, non-relativistic to relativistic---covering the full spectrum of feasible computational architectures. In so doing, a universal scaling limit is discovered, exceeding that of irreversible computers yet still falling short of the volumetric quantum bound.

\section{Quantum Ballistic Architectures and the Quantum Zeno Effect}
\label{sec:qze}

We first consider a quantum architecture. A viable quantum architecture must be ballistic in the sense of following an exact prescribed trajectory in phase space, as to do otherwise would lead to decoherence undermining the computational state. Of course we have established that a ballistic quantum system cannot truly be isolated from entropic effects; in particular, the third law of thermodynamics precludes lowering the temperature of any system to absolute zero, and so we must incorporate a heat bath into our analysis.

Suppose the ballistic Hamiltonian is given by $H_0$ and the perturbative effect of the heat bath by $V(t)$, such that the true Hamiltonian is $H=H_0+V$. Further, let the initial density be $\rho$ and introduce the projector $P$ corresponding to the subspace of valid states, such that $P\rho=\rho P=\rho$. To ensure processive computation, we must therefore periodically correct errors introduced by the perturbation. The cost of such error correction will be bounded from below by the entropy increase due to $V$, which can be expressed in terms of the probability of an error $\delta p$ corresponding to the erroneous subspace $P^\perp=1-P$.

If corrections are to be made at intervals $\delta t$, then this error will be found to be $\delta p(t)=\tr[P^\perp \rho(t+\delta t)]$.
For a time-dependent Hamiltonian, the quantum Liouville equation tells us that $\dot\rho=i[\rho,H/\hbar_P]$ where $\hbar_P$ is Planck's constant.
We can expand $\rho$ as
\begin{align*}
  \rho(t+\delta t) &= \rho + \delta t \pdv{\rho}{t} + \tfrac12\delta t^2 \pdv[2]{\rho}{t} + \bigO{\delta t^3} \\
  &= \rho + i[\rho, \varepsilon] +
    \tfrac12( i\delta t[\rho,\dot\varepsilon] - (\varepsilon^2\rho - 2\varepsilon\rho\varepsilon + \rho\varepsilon^2)) + \bigO{\delta t^3}
\end{align*}
where we have omitted explicit time dependence for brevity and written $\varepsilon\equiv H\delta t/\hbar_P$. We then find
\begin{align*}
  \delta p &= \tr[P^\perp (\rho + i[\rho,\varepsilon + \tfrac12\delta t\dot\varepsilon] - \tfrac12(\varepsilon^2\rho - 2\varepsilon\rho\varepsilon + \rho\varepsilon^2))] + \bigO{\varepsilon^3} \\
    &= \tr[P^\perp \varepsilon\rho\varepsilon] + \bigO{\varepsilon^3}
       \equiv \tr[\rho\varepsilon P^\perp \varepsilon] + \bigO{\varepsilon^3}
\end{align*}
where the last line uses the fact that $P^\perp\rho=\rho P^\perp=0$. Note that if the $\rho\varepsilon P^\perp\varepsilon$ term vanishes, then so do all higher terms, and therefore this term is always the leading order approximant.

In order to evaluate this trace, we first write $H_0$ and $V$ as block matrices in the basis $(P,P^\perp)$, yielding
\begin{align*}
  H_0 &= \begin{pmatrix}
    h_{00} & 0 \\ 0 & h_{11}
  \end{pmatrix}\,, &
  V &= \begin{pmatrix}
    v_{00} & v_{01} \\ v_{10} & v_{11}
  \end{pmatrix}\,.
\end{align*}
Using the idempotence of projectors, i.e.\ $P^\perp\equiv P^\perp P^\perp$, and the fact $\rho\equiv P\rho$, we can see that $\rho HP^\perp H\equiv \rho VP^\perp V=\rho(V^2-VPV)$. Therefore, we have
\begin{align*}
  \delta p &= \qty\Big(\frac{\delta t}{\hbar_P})^2 (\evQty{V^2} - \evQty{VPV})\,.
\end{align*}
The first thing to notice about this expression is that it varies with the square of $\delta t$. This is characteristic of the Quantum Zeno Effect (QZE~\cite{qze})---that is, in the limit of constant measurement (i.e.\ $\delta t\to 0$), we can freeze dynamical evolution. Here, we are performing a partial measurement by projecting onto the subspace of either valid or erroneous computational states, thus allowing computational evolution to continue. Of course, the Zeno rate itself is subject to the Margolus-Levitin limit~\cite{margolus-levitin} and so such constant measurement is not possible.

To proceed in further determining $\delta p$, and thereby $\delta h$, it is important to elaborate on certain architectural details. Margolus and Levitin's analysis shows that one gets the same $E/h_P$ total rate regardless of whether one considers transitions of the system as a whole (`serial') or the combined transitions of a system partitioned into subsystems (`parallel'). For a physically realistic system, measuring large subsystems as a whole is impractical and so a more fine-grained architecture is likely preferable. Nevertheless, we shall proceed generally, finding the result is independent of this detail.

Let the subsystems be indexed by $i\in\mathcal I$. For a fully serial system, $|\mathcal I|=1$. We allow each subsystem to evolve independently, assuming that interaction events are negligibly frequent relative to our Zeno corrections. Specialising our above result for $\delta p$, we have
\begin{align*}
  \delta p_i &= \qty\Big(\frac{\delta t_i}{\hbar_P})^2 (\evQty{V_i^2}-\evQty{V_iP_iV_i})
\end{align*}
The strength of the perturbations $V_i$ will be controlled by an effective temperature for that subsystem, $T_i$. Suppose the subsystem has $n_i$ degrees of freedom\footnote{Excluding any frozen out by low temperatures.}, hereafter referred to as \textit{computational primitives}. By the equipartition theorem, the energies of each primitive will be individually perturbed. For an uncorrelated perturbation, we expect that $\overline{\evQty{V}}=0$ where $\overline x$ is the time-average value of $x$. Therefore, $\overline{\evQty{V^2}}\equiv\var_{\evQty{\bar\cdot}} V$. The expected aggregate perturbation over all primitives, then, is $n_i (k T_i)^2$ assuming $V$ is Gaussian and where $k\equiv k_B$ is Boltzmann's constant.

As for the second term, $VPV$, we expect that on average $V$ will mix (near-)degenerate states indiscriminately. $VPV$ represents the probability that the perturbation stays within the intended computational subspace. The state space can be divided into disjoint computational subspaces, within which the dynamics perform a reversible cycle over computational states. Let the total state space have cardinality $\Omega_i$, and the number of microstates associated with each subspace be $\omega_{ij}$ where $j$ indexes the different subspaces such that $\Omega_i=\sum_j\omega_{ij}$. Assuming similar energy distributions between subspaces, the chance of remaining within the original subspace is $\omega_{ij}/\Omega_i$. We further approximate this by $\omega_i/\Omega_i$ where $\omega_i=\evQty{\omega_{ij}}_j$ is the average subspace cardinality. Equivalently, we can define $\Omega_i/\omega_i \equiv W_i$, the number of distinct programs that the subsystem is capable of executing. Therefore,
\begin{align*}
  \overline{\delta p_i} &= n_i \qty\Big(\delta t_i \frac{k T_i}{\hbar_P})^2 (1-\sfrac1{W_i})
\end{align*}
In fact, we may wish to correct the second term errors in which we jump within the same subspace, as this may interfere with synchronisation between different subsystems (or it may be difficult to ascertain whether we have jumped within the same subspace). In this case, the expression reduces to
\begin{align*}
  \overline{\delta p_i} &= n_i \qty\Big(\delta t_i \frac{k T_i}{\hbar_P})^2 (1-\sfrac1{\Omega_i})
\end{align*}
In any case, $1-\sfrac1{W_i}$ is always finite and $\in[\tfrac12,1)$ for any useful subsystem; that is, a useful subsystem should have $W_i\ge2$. For composite/non-primitive subsystems, we would like the number of useful programs to scale exponentially with the number of primitives, i.e.\ $W_i=g_i^{n_i}$ for some $g_i$ (typically on the order of unity but greater than 1). For a composite subsystem of even moderate size, this exponential scaling will render the $1-\sfrac1{W_i}$ term effectively unity.

We are now able to determine the entropy increase between Zeno cycles. The indiscriminate mixing of the perturbation means that, in the event of an error, the entropy attains its maximum value. This yields the following expression for the information entropy\footnote{We use $H$ to refer to the information theoretical entropy, connected to the thermodynamical entropy as $S=k_BH$. Furthermore, we use the lowercase quantity $h$ to refer to the entropy of a subsystem or particle.},
\begin{align*}
  \delta h_i &= [-(1-\delta p_i)\log(1-\delta p_i) - \delta p_i\log\delta p_i
      + (1-\delta p_i)\log\omega_i + \delta p_i\log\Omega_i] - [\log\omega_i] \\
      &= \delta p_i (1 + \log\tfrac{\Omega_i}{\omega_i} - \log\delta p_i) - \bigO{\delta p_i^2} \\
      &\equiv \delta p_i (\log W_i + 1 - \log\delta p_i) - \bigO{\delta p_i^2} \\
      &\ge n_i\delta p_i \log g_i - \bigO{\delta p_i^2}\,.
\end{align*}
This expression assumes ignorance of states within the current computational subspace; again, we may wish to correct these errors too, in which case we may simply substitute $\omega_i=1$. This would effectively lead to taking $W_i=\Omega_i=g_i'^{n_i}$ where $g_i'\ge g_i$, and so the inequality remains valid.

Putting these together, we get
\begin{align*}
  \dot h_i &\ge \frac{n_i^2}{r_{Z,i}} \underbrace{\zeta_i\qty\Big(\frac{kT_i}{\hbar_P})^2 \log g_i}_{\equiv\gamma_i}
\end{align*}
where $r_{Z,i} \equiv 1/\delta t_i$ is the Zeno measurement rate for the subsystem and $\zeta_i=1-\sfrac1{W_i}\in[\tfrac12,1)$. The aggregate rate of entropy generation is given by $\dot H = \sum_i \dot h_i$ and is subject to the constraint $k\hat T\dot H\le P$ where $\hat T$ is the system temperature that governs the Landauer bound and $P=\phi A$ is the heat dissipation power, proportional to the system's surface area. It may seem surprising that $\dot h_i\propto n_i^2$. In fact, this only applies for sufficiently small $\delta p_i$; when $\delta p_i$ becomes significantly large, $\dot h_i$ approaches its maximum of $n_i r_{Z,i} \log g_i$.

We wish to maximise the computational rate, subject to this constrained entropy production. Per Margolus and Levitin, this rate depends on the combined energy of the computational primitives. If the average energy per primitive in subsystem $i$ is $\varepsilon_i$, then the energy available for computation is $E_C=\sum_i \varepsilon_i n_i$ and the computational rate is subject to $R_C\le E_C/h_P$.

Introducing the Lagrangian multiplier $\alpha$, we wish to maximise $\Lambda$ with respect to $n_i$;
\begin{align*}
  \Lambda = \sum_i\varepsilon_i n_i - \frac{1}{2\alpha} \sum_i\frac{n_i^2}{r_{Z,i}} \gamma_i 
  \qquad\implies\qquad
  0 = \varepsilon_i - \frac1\alpha \frac{n_i}{r_{Z,i}} \gamma_i\,.
\end{align*}
This gives $\dot h_i \ge \alpha \varepsilon_i n_i$ and so
\begin{align*}
  \dot H \ge \alpha E_C\,;
\end{align*}
solving for $\alpha$ and summing over $i$,
\begin{align*}
  \alpha r_{Z,i} &= \frac{n_i \gamma_i}{\varepsilon_i} \equiv \frac{n_i\varepsilon_i \gamma_i}{E_C\varepsilon_i^2} E_C \\
  \alpha R_Z &= E_C \evqty\Big{\frac{\gamma_i}{\varepsilon_i^2}}_{E_{C,i}}\,.
\end{align*}
where the average is taken over the computational energy distribution and $R_Z$ gives the net Zeno rate of the system as a whole. Substituting into the $\dot H$ constraint and making use of the fact that $R_Z\le E_Z/h_P$,
\begin{align*}
  \frac{P}{k\hat T} \ge \dot H &\ge \frac{E_C^2}{R_Z} \evqty\Big{\frac{\gamma}{\varepsilon^2}} \\
  \frac{P}{k\hat T}\frac{E_Z}{h_P} \evqty\Big{\frac{\gamma}{\varepsilon^2}}^{-1} &\ge E_C^2\,
\end{align*}
where $\hat T$ can be called the `Szilard' temperature: the temperature of the system in which entropy is generated and must be subsequently erased.
Finally, we obtain an expression for the net computational rate $R_C$,
\begin{align*}
  R_C &\le \sqrt{\frac{P}{k\hat T} \frac{E_Z}{h_P} \evqty\bigg{\zeta_i\qty\bigg(\frac{2\pi kT}{\varepsilon})^2 \log g}^{-1}}
  = \frac1{2\pi} \evqty\bigg{\frac{\zeta_i\log g}{\beta^2\varepsilon^2}}^{-1/2} \sqrt{\frac{P}{k\hat T}\frac{E_Z}{h_P}}\,.
\end{align*}
Now, as the upper bound of $P$ is proportional to the bounding surface area, and the upper bound of $E_Z$ is proportional to the system's volume, we get the scaling law
\begin{align*}
  R_C &\lesssim \sqrt{AV} \sim V^{5/6}\,,
\end{align*}
consistent with the scaling law for adiabatic architectures found by \textcite{frank-thesis} and showing it to be an upper bound.

We note that \textcite{levitin-toffoli-full} derive a related expression for the rate of energy dissipation/heat production within a quantum harmonic oscillator (QHO),
\begin{align*}
  P &= \frac{\varepsilon h_P R_C^2}{(1-\varepsilon)N}\,,
\end{align*}
where $\varepsilon$ is the probability of error per state transition and $N$ is the number of QHO energy levels. Given that their analysis assumes maximal $R_C$, however, this simplifies to
\begin{align*}
  P &= \frac{\varepsilon\Delta E}{1-\varepsilon} R_C
\end{align*}
where $\Delta E$ is the separation of energy levels. Given that $\varepsilon$ is assumed constant in the analysis, this then yields an areametric bound on $R_C$. Our analysis permits $\varepsilon$ to vary, and its optimum value as a function of system parameters is obtained. We also remain general in the systems studied, and thus confirm the conjecture that for a fixed error probability the energy-dissipation rate increases quadratically with the rate of computation.

We also note the recent results\footnote{The author gratefully acknowledges Mike Frank for pointing him towards this body of work.} of \textcite{pidaparthi-lent} on the Landau-Zener effect (LZE)~\cite{lze1,lze2}. In a closed quantum system, with no thermal coupling to the external environment, the Landau-Zener effect predicts that the energy dissipation decays \emph{exponentially} with the switching time of the system (the inverse of the computation rate). That this is non-zero even in the absence of an external environment shows the difficult in achieving ballistic computation even in `idealised' conditions, but it certainly points towards a route to achieving near-ballistic computation in regimes in which thermal coupling is negligible. As expected, once thermal coupling is re-introduced to the system Pidaparthi and Lent show how the adiabatic behaviour is recovered. Moreover, their numerical results indicate that the LZE does not appear to be a practicable approach to achieving super-adiabaticity, except perhaps in a very narrow region of system conditions and size. Moreover, the results of Pidaparthi and Lent show the occurrence of a local minimum of dissipation with respect to switching time that would lead to engineering difficulties in increasing computer size for this parameter range. In any case, our asymptotic results stand, although it would be instructive to characterise the parameter range in which the LZE is exploitable and in which the aforementioned engineering difficulties arise.

\section{Classical Architectures I: A General Lagrangian Approach}

If use of the QZE is not possible, perhaps because of high temperatures or an excessive Zeno rate, then Fermi's golden rule applies, giving $\delta p\propto\delta t$. Consequently $\dot h$---and thus $\dot H$---is subject to a finite lower bound. We find $\dot h\ge n\dot p\log g$, yielding $\dot H\ge E_C\evqty{\dot p\log g/\varepsilon}$. Combining with our constraint $\dot H\le\hat\beta P$, the inequality in $R_C$ becomes
\begin{align*}
  R_C &\le \evqty{\frac{\dot p\log g}{\varepsilon/h_P}}^{-1}\frac{P}{k\hat T}
\end{align*}
where $h_P$ here is Planck's constant, and so the scaling law falls to $R_C\lesssim A\sim V^{2/3}$. That is, any ballistic system not making use of the QZE will be subject to the same areametric limit as irreversible computers.

The breakdown of the QZE corresponds to an increase in thermal coupling. By abandoning the desire to maintain well defined quantum states, we instead enter the incoherent regime of the classical realm wherein quantum effects are significantly suppressed. Proceeding generally, we adopt a Lagrangian formalism in order to remain noncommittal in our choice of coordinates $\vec q$.

We introduce a reasonably general Lagrangian, $\mathcal L=T-V$,
\begin{align*}
  \mathcal L &= [\tfrac12 c_{ij}\dot q_i\dot q_j] - [V + w_i\dot q_i + \mathcal O(\dot{\vec q}{\,}^3)] \equiv \tfrac12\dot q^\top C\dot q - V - W\dot q + \mathcal O(\dot q^3)
\end{align*}
where the coefficients $c$, $V$, $w$...\ may depend continuously on the coordinates $\vec q$, and Einstein notation is used for repeated indices. We have rewritten the Lagrangian in matrix form for concision, where $C$ is a matrix, $V$ a scalar, $W$ a row vector, and $q$ a column vector. We assume the system is (effectively) closed, self-contained, or otherwise independent of its environment, and thus $\mathcal L$ will not formally depend on the time parameter. In addition, this property implies invariance under global translation of any of its coordinates and thus conservation of all associated momenta. We obtain the Hamiltonian $\mathcal H$ and generalised momenta thus
\begin{align*}
  \mathcal H &= \pdv{\mathcal L}{\dot q_i}\dot q_i - \mathcal L = \tfrac12\dot q^\top C\dot q + V + \mathcal O(\dot q^3) \\
  p_i &= \pdv{\mathcal L}{\dot q_i} = C\dot q - W + \mathcal O(\dot q^3)
\end{align*}
Notice that the terms of our potential linearly dependent on the velocities drop out of the Hamiltonian. Those dependent on the cube or higher remain, but will turn out to be negligible at the velocities optimal for computation (and are typically not physically relevant in any case).

For brief collisions, the change in coordinates is negligible and so the spatiotemporal variance in coefficients can be neglected. In this case, we should conserve the Hamiltonian and momenta. We introduce superscripts to the coordinates, $q_i^{(n)}$, to represent different particles. We also introduce notation for the velocities before, $u$, after, $v$, and their change, $\Delta u=v-u$. For a collision between particles $m$ and $n$,
\begin{align*}
  \Delta\mathcal H &= [\tfrac12 v^{(m)\top}C^{(m)}v^{(m)} - \tfrac12 u^{(m)\top}C^{(m)}u^{(m)}] + [\tfrac12 v^{(n)\top}C^{(n)}v^{(n)} - \tfrac12 u^{(n)\top}C^{(n)}u^{(n)}] \\
    &= \tfrac12 (u^{(m)} + v^{(m)})^\top C^{(m)}\Delta u^{(m)} + \tfrac12 (u^{(n)} + v^{(n)})^\top C^{(n)}\Delta u^{(n)} = 0 \\
  \Delta p &= C^{(m)}\Delta u^{(m)} + C^{(n)}\Delta u^{(n)} = 0\,,
\end{align*}
where we use the fact that $C$ is symmetric. Substituting $\Delta p$ into $\Delta\mathcal H$, we find,
\begin{align*}
  2\Delta\mathcal H^{(m,n)} &= [(u^{(m)}+v^{(m)})-(u^{(n)}+v^{(n)})]^\top C^{(m)}\Delta u^{(m)} \\
  0 &= [(2u^{(m)}+\Delta u^{(m)})-(2u^{(n)}+\Delta u^{(n)})]^\top C^{(m)}\Delta u^{(m)} \\
    &= [2(u^{(m)}-u^{(n)}) + (1+C^{(n)-1}C^{(m)})\Delta u^{(m)}]^\top C^{(m)}\Delta u^{(m)} \\
    &= [2(u^{(m)}-u^{(n)}) + (C^{(m)-1}+C^{(n)-1})C^{(m)}\Delta u^{(m)}]^\top C^{(m)}\Delta u^{(m)} \\
    &\equiv [2(u^{(m)}-u^{(n)}) + \mu^{(m,n)-1}C^{(m)}\Delta u^{(m)}]^\top C^{(m)}\Delta u^{(m)} \\
    &= [2\underbrace{\mu^{(m,n)+\sfrac12}(u^{(m)}-u^{(n)})}_{-y} + \underbrace{\mu^{(m,n)-\sfrac12}C^{(m)}\Delta u^{(m)}}_{x}]^\top \mu^{(m,n)-\sfrac12}C^{(m)}\Delta u^{(m)}\,,
\end{align*}
where $\mu$ is the generalised reduced mass of the $(m,n)$ system. We can solve for $x$, and thereby $\Delta u^{(m)}$, by rewriting thus,
\begin{align*}
  |x-y|^2 &= |y|^2\,.
\end{align*}
This form reflects the fact that we lack sufficient constraints to unambiguously solve for the post-collision picture. In Cartesian coordinates for point particles, this manifests as an unknown direction of motion afterwards. This is usually resolved by introducing the constraint that momenta perpendicular to the normal vector between the colliding bodies are unchanged. In our generalised coordinate system, the appropriate constraint is unspecified. The general solution is given by
\begin{align*}
  x &= y+|y|\hat n = |y|(\hat y + \hat n)
\end{align*}
where $\hat n$ is a unit vector of unknown direction. We now find an expression for the change in kinetic energy of our particle of interest, $(m)$,
\begin{align*}
  \Delta\mathcal H^{(m)} &= (u^{(m)} + \tfrac12\Delta u^{(m)})^\top C^{(m)}\Delta u^{(m)} \\
    &= u^{(m)\top} \mu^{(m,n)\sfrac12}|y|(\hat y+\hat n) + \bigOO{\Delta u^{(m)2}} \\
    &= \eta|u^{(m)\top}\mu^{(m,n)}(u^{(m)}-u^{(n)})| + \bigOO{\Delta u^{(m)2}}
\end{align*}
where we have introduced a factor $\eta\in[-2,2]$ to take into account our uncertainty in the final direction, and where we have assumed $|\Delta u|\ll u$. Being more careful, we can determine $\Delta u^{(m)}$ thus,
\begin{align*}
  \Delta u^{(m)} &= C^{(m)-1} \mu^{(m,n)+\sfrac12} |\mu^{(m,n)+\sfrac12}(u^{(n)}-u^{(m)})| (\hat y+\hat n) \\
  |\Delta u^{(m)}| &= \eta' |C^{(m)-1} \mu^{(m,n)}(u^{(n)}-u^{(m)})| \\
    &= \eta' |(C^{(m)}+C^{(n)})^{-1} (C^{(n)} u^{(n)} + C^{(m)} u^{(m)}) - u^{(m)}| \\
    &= \eta' |\tilde u^{(m,n)} - u^{(m)}|
\end{align*}
where $\tilde u^{(m,n)}$ is the average of prior velocities, weighted by their generalised masses $C^{(\cdot)}$, and $\eta'\in[0,2]$ is another uncertainty factor related to $\eta$. When $u^{(m)}$ is large compared to $u^{(n)}$, $(u^{(m)}+\Delta u^{(m)})\approx u^{(m)}$. When it is small, $(u^{(m)}+\Delta u^{(m)})\approx u^{(n)}$. Thus a better approximation to $\Delta\mathcal H$ is given by
\begin{align*}
  \Delta\mathcal H^{(m)} &\approx \eta|\tilde u^{(m,n)\top}\mu^{(m,n)}(u^{(m)}-u^{(n)})|\,.
\end{align*}

$\Delta\mathcal H$ gives the energy lost from the computational particle $(m)$ to some environmental particle $(n)$, and thus is the amount of energy that must be supplied to the computational system and dissipated from the thermal system. In order to proceed we must determine the rate of this energy transfer; if we assume that the kinetic dynamics of the system are significantly faster than the computational transitions, as suggested by our inability to sufficiently control the quantum state, then we can assume negligible extant correlations between the positions and velocities of different particles. Thus we can employ a kinetic theory approach to determine this rate.

We first determine the mean free path $\ell$, the average distance a particle travels between collisions. Given the vanishing correlations, we are able to treat collisions between different species of particle separately. For spherically symmetric particles, we find $\pi (r^{(\alpha)} + r^{(\beta)})^2 \ell^{(\alpha;\beta)} = 1/n^{(\beta)}$ for the mean free path of $\alpha$ particles between collisions with $\beta$ particles, where $n^{(\beta)}=N^{(\beta)}/V^{(\beta)}$ is the number density of $\beta$ particles. More generally, we can replace $\pi(r^{(\alpha)}+r^{(\beta)})2$ with $A^{(\alpha,\beta)}$, the effective collision cross-section for arbitrarily shaped particles averaged over impact orientation. We also need to determine the average relative speed between $\alpha$ and $\beta$ particles,
\begin{align*}
  \bar v_{\text{rel}}^{(\alpha,\beta)2} &= \evQty{(v^{(\alpha)} - v^{(\beta)})^2} \\
    &= \evQty{v^{(\alpha)2}} + \evQty{v^{(\beta)2}} - 2\evQty{v^{(\alpha)} \cdot v^{(\beta)}} \,,\\
  \bar v_{\text{rel}}^{(\alpha,\beta)} &= \sqrt{\bar v^{(\alpha)2} + \bar v^{(\beta)2}} \,,\\
  \nu^{(\alpha;\beta)} &= \bar v_{\text{rel}}^{(\alpha,\beta)} / \ell^{(\alpha;\beta)} \\
    &= A^{(\alpha,\beta)}n^{(\beta)}\sqrt{\bar v^{(\alpha)2} + \bar v^{(\beta)2}}\,,
\end{align*}
where we have used the fact that the velocities are uncorrelated to cancel the cross term. Summing over these rates for each class, we can find the rate of loss of kinetic energy for $\alpha$ particles,
\begin{align*}
  \dot{\mathcal H}^{(\alpha)} &= \sum_\beta \eta^{(\alpha;\beta)} |\tilde u^{(\alpha,\beta)\top} \mu^{(\alpha,\beta)} (u^{(\beta)} - u^{(\alpha)})| A^{(\alpha,\beta)}n^{(\beta)} \sqrt{\bar u^{(\alpha)2} + \bar u^{(\beta)2}}\,.
\end{align*}
When $\alpha$ particles are heavy and fast, this reduces to
\begin{align*}
  \dot{\mathcal H}^{(\alpha)} &= \sum_\beta \eta^{(\alpha;\beta)} |u^{(\alpha)\top} C^{(\beta)} u^{(\alpha)}| A^{(\alpha)}n^{(\beta)} \bar u^{(\alpha)} \\
    &= \sum_\beta \eta^{(\alpha;\beta)} \rho^{(\beta)} A^{(\alpha)} \bar u^{(\alpha)3}
\end{align*}
where $\rho=n|C|$ is the generalised density. Notice that this takes the same form as the equation for hydrodynamic drag, taking the drag coefficient to be $2\eta$. Assuming there exists some correspondence between the generalised velocity $u^{(\alpha)}$ and the rate of computation, we find
\begin{align*}
  P &\ge \sum_\alpha N^{(\alpha)} A^{(\alpha)}\qty\bigg(\ell^{(\alpha)}\frac{R^{(\alpha)}}{N^{(\alpha)}})^3 \sum_\beta \eta^{(\alpha;\beta)}\rho^{(\beta)}
\end{align*}
where $\alpha$ ranges over computational particles and $\ell$ is the characteristic generalised displacement corresponding to a single computational transition. To maximise the net computational rate $R=\sum_\alpha R^{(\alpha)}$, we find that we need to let $\bar u^{(\alpha)}\to 0$, but this violates our assumption that $\alpha$ particles are fast. For finite velocity computational particles, we thus find that out net computational rate scales as $R\lesssim A\sim V^{2/3}$.

In the limit of slow $\alpha$ particles, $|u^{(\alpha)}|\ll|u^{(\beta)}|$, $\dot{\mathcal H}$ instead reduces to
\begin{align*}
  \dot{\mathcal H}^{(\alpha)} &= \sum_\beta \eta^{(\alpha;\beta)} |\tilde u^{(\alpha,\beta)\top}C^{(\beta)}u^{(\alpha)}|A^{(\alpha)}n^{(\beta)}\bar u^{(\beta)}
\end{align*}
where we have taken $\evQty{u^{(\beta)}-u^{(\alpha)}}=\bar u^{(\alpha)}$. If the generalised mass of the $\alpha$ particles is not significantly heavier than the $\beta$ particles, then $\tilde u^{(a,b)}$ will have a strong dependence on $u^{(\beta)}$ and this will lead to the same scaling limit of $R\lesssim A$. In order to improve on this, we must let the $\alpha$ particles be significantly heavier. In this limit, $\tilde u^{(\alpha,\beta)}\approx u^{(\alpha)}$ and we get
\begin{align*}
  \dot{\mathcal H}^{(\alpha)} &= \sum_\beta \eta^{(\alpha;\beta)} |u^{(\alpha)\top}C^{(\beta)}u^{(\alpha)}|A^{(\alpha)}n^{(\beta)}\bar u^{(\beta)} \,,\\
  P &\ge \sum_\alpha N^{(\alpha)}A^{(\alpha)}\qty\bigg(\ell^{(\alpha)}\frac{R^{(\alpha)}}{N^{(\alpha)}})^2 \sum_\beta \eta^{(\alpha;\beta)} \rho^{(\beta)}\bar u^{(\beta)} \\
    &\ge \sum_\alpha \frac{R^{(\alpha)2}}{N^{(\alpha)}} \underbrace{A^{(\alpha)}\ell^{(\alpha)2} \sum_\beta \eta^{(\alpha;\beta)} \rho^{(\beta)}\bar u^{(\beta)}}_{\gamma^{(\alpha)}}\,.
\end{align*}
To proceed, we maximise the rate subject to this power constraint,
\begin{align*}
  \Lambda = \sum_\alpha R^{(\alpha)} - \frac{1}{2\lambda}\sum_\alpha \frac{R^{(\alpha)2}\gamma^{(\alpha)}}{N^{(\alpha)}}
  \qquad\implies\qquad
  0 = 1 - \frac{R^{(\alpha)}\gamma^{(\alpha)}}{\lambda N^{(\alpha)}}
\end{align*}\begin{align*}
  \frac R\lambda &= \sum_\alpha \frac{n^{(\alpha)}}{N\gamma^{(\alpha)}}N = N\evqty\bigg{\frac1{\gamma^{(\alpha)}}} \\
  P &\ge \sum_\alpha \lambda R^{(\alpha)} = \lambda R = \frac{R^2}{N}\evqty\bigg{\frac1{\gamma^{(\alpha)}}}^{-1} \\
  R &\le \sqrt{PN\evqty\bigg{\frac1\gamma}}
\end{align*}
where $N=\sum_\alpha N^{(\alpha)}$ is the total number of computational particles. To maximise the computational rate $R$, then, we let $N$ scale with the volume of the system, leading once again to the scaling law
\begin{align*}
  R \lesssim \sqrt{AV} \sim V^{5/6}\,.
\end{align*}

\section{Classical Architectures II: Brownian Machines}
\label{sec:crn}

Whilst the previous two sections should suffice to cover all physical computational systems, the high mass-low speed limit of the classical system is not ideal from an engineering perspective. For improved practicality, we reformulate this limit in terms of an abstract chemical reaction network (CRN) near equilibrium, and show that we obtain the same result. This result is thus more robust in terms of its attainability.

\paragraph{Entropy generation rate}
We first seek a general expression for the rate of entropy generation for a chemical reaction network. Consider any such dynamical system consisting of a set of species $\{C_j:j\}$ and a set of reversible reactions $\Gamma = \{ \nu_{ij}X_j \longleftrightarrow_{\gamma_i} \nu_{ij}'X_j : i \}$, where we sum over $j$ and the $\nu_{ij}$ are stoichiometries. By the convergence theorem for reversible Markov systems, the system will converge to a unique steady state/equilibrium distribution for any initial conditions. As the reactions are reversible, the steady state will also satisfy detailed balance such that the forward and backward rates coincide separately for each reaction, i.e.\ $\smfwd R_i=\smbwd R_i$, where $R$ is a rate of the whole system rather than per unit volume.

Let us look for a quantity $H$ that measures progress towards equilibrium and satisfies the following properties:
\begin{enumerate}
  \item $H$ is a state function of the system\footnote{More accurately, on the joint state of an ensemble of iid systems.}, depending only on its instantaneous description;
  \item $H$ increases monotonically with time;
  \item $H$ is additive, i.e.\ $H(\cup_iV_i)=\sum_iH(V_i)$ for any set of disjoint regions $V_i$.
\end{enumerate}
By properties 1 and 2 and the convergence theorem, $H$ will approach a unique maximum at equilibrium. Microscopically, reactions occur discretely and so the rate of change of $H$ can be written
\begin{align*}
  \dot H_i &= \fwd R_i\fwd{\Delta h}_i+\bwd R_i\bwd{\Delta h}_i = (\fwd R_i-\bwd R_i)(h_{\text{reactants}}-h_{\text{products}})
\end{align*}
for any reaction $i$, where we have used reversibility to identify $\smfwd{\Delta h}_i=-\smbwd{\Delta h}_i=h_{\text{reactants}}-h_{\text{products}}$ and where $h(\sum_j\nu_{ij}X_j)$ is the entropy of the region associated with precisely $\nu_{ij}$ particles of each species $X_j$. In order to satisfy property 2, the sign of $\smfwd{\Delta h}_i$ must equal that of $\smfwd R_i-\smbwd R_i$.

Now, consider fixing $h_{\text{products}}$ and $\smbwd R_i$, and varying $\smfwd R_i$; this is possible unless the reaction is trivial with $\nu_{ij}=\nu_{ij}'$ for all species $X_j$, i.e.\ it does nothing. To maintain property 2, it must therefore be the case that $\smfwd{\Delta h}_i=h(\smfwd R_i)-h(\smbwd R_i)$. To satisfy property 3, we require $h(\alpha\smfwd R_i)-h(\alpha\smbwd R_i)=h(\smfwd R_i)-h(\smbwd R_i)$ where $\alpha$ is some arbitrary scaling factor of the system. We can therefore infer the functional form $h(x)=\log_b ax$ for some constants $a$ and $b$.

We have therefore derived an entropy like quantity, unique up to choice of logarithmic unit via $b$ and entropy at $T=0$ via $a$. Without loss of generality we choose $a=1$ and $b=e$, and therefore find that
\begin{align*}
  \dot H &= \sum_i (\fwd R_i - \bwd R_i)\log\frac{\fwd R_i}{\bwd R_i} \equiv 2\sum_i R_i\beta_i\arctanh \beta_i
\end{align*}
where $\beta_i=(\smfwd R_i-\smbwd R_i)/(\smfwd R_i+\smbwd R_i)$ is the effective bias of reaction $i$ and $R_i=\smfwd R_i+\smbwd R_i$ is its gross reaction rate.

\paragraph{Entropy generation rate for elementary chemical reactions}

We now calculate our entropy quantity for elementary chemical reactions and compare against the expected value. The rates for elementary chemical reactions are given by collision theory as
\begin{align*}
  \fwd r_i &= \fwd k_i \prod_j[X_j]^{\fwd\nu_{ij}} &
  \bwd r_i &= \bwd k_i \prod_j[X_j]^{\bwd\nu_{ij}}
\end{align*}
where $\smfwd k_i$ and $\smbwd k_i$ are rate constants and $r$ are rates per unit volume. Assuming no inter-particle interactions, the canonical entropy is given by the Sackur-Tetrode equation which can be obtained simply by considering the spatial distribution of the particles along with any other degrees of freedom. For species $X_i$, consider an arbitrary volume $V_i$ available to it; if the thermal volume of the particles is $\mathcal V_i=\Lambda^3$ where $\Lambda$ is the thermal \emph{de Broglie} wavelength, then there will be $V_i/\mathcal V_i$ loci available to the particle, and so the associated entropy within the volume will be
\begin{align*}
  N_ih(X_i) &= N_i \log\frac{V_i}{\mathcal V_i} - \underbrace{\log N_i!}_{\mathclap{\text{Gibb's factor}}} + N_i(\varepsilon_i-1) \\
  h(X_i) &= \varepsilon_i - \log[X_i]\mathcal V_i + \bigO{\frac{\log N_i}{N_i}}
\end{align*}
where $N_i=[X_i]V_i$ is the number of particles in the volume, and $[X_i]$ the concentration. We have also taken into account indistinguishability of the particles with the Gibb's factor, and allowed for additional degrees of freedom with the $\varepsilon_i$ term. Strictly speaking the spatial distribution should be calculated combinatorially via the multinomial coefficient
\begin{align*}
  \log\frac{(V/\mathcal V)!}{((V/\mathcal V - \sum N_i))!\prod N_i!}
\end{align*}
where we have assumed all the particles inhabit the same region and have the same wavelength. Applying Stirling's approximation and assuming that $\sum N_i\ll V/\mathcal V$, this reduces to $\sum N_i(1-\log[X_i]\mathcal V)$ however, so our simplistic derivation is valid.

Therefore we find the canonical entropy change due to reaction $i$ is given by
\begin{align*}
  \Delta h_i &= \sum_j(\bwd\nu_{ij}-\fwd\nu_{ij})h(X_j)
     = \sum_j(\bwd\nu_{ij}-\fwd\nu_{ij})(\varepsilon_j-\log\mathcal V_j) - \sum_j(\bwd\nu_{ij}-\fwd\nu_{ij})\log[X_j]
\end{align*}
using a slight abuse of notation (the logarithmands are not unitless, but their combination is). Comparing with our quantity $\Delta h=\log\smfwd r/\smbwd r$, we find
\begin{align*}
  \Delta h_i &= \log\frac{\fwd k_i}{\bwd k_i} - \sum_j(\bwd\nu_{ij}-\fwd\nu_{ij})\log[X_j] \\
  \frac{\fwd k_i}{\bwd k_i} &= \qty\bigg(\prod_j\mathcal V_j^{\fwd\nu_{ij}-\bwd\nu_{ij}})\qty\bigg(\exp\sum_j(\fwd\nu_{ij}-\bwd\nu_{ij})\varepsilon_j)
\end{align*}
which should be compared to the Arrhenius equation. Indeed, the $\mathcal V_i$ terms confer the appropriate units to the pre-exponential factor and, by expanding the enthalpy term in terms of $k_BT$ as $\varepsilon_i=\varepsilon_i^{(0)}+\frac1{k_BT}\varepsilon_i^{(1)}+\bigOO{\frac1{k_BT}}^2$, we can group the roughly constant terms of the enthalpy into the pre-exponential factor, leaving the exponential term in the form $e^{-\Delta E/k_BT}$.

\paragraph{Maximising performance in Brownian machines}

We wish to maximise the net reaction rate of the system by selecting the biases $\beta_i$ of each individual reaction, subject to bounded total entropy production. If $R=\sum R_i$ is the total gross reaction rate, then the total net rate is $R_c=\sum R_i\beta_i=R\evqty{\beta}$ where the expectation value is with respect to the fractional contribution of each reaction, i.e.\ weighted by $R_i/R$. We denote this by $R_c$ assuming that each reaction contributes to computational progress; if this is not true then $R_c$ will be less than this value. The entropy rate is similarly given by $\dot H = 2R\evqty{\beta\arctanh\beta}$. We use a Lagrange multiplier approach as usual,
\begin{align*}
  \Lambda &= R\evqty{\beta} - \lambda R\evqty{\beta\arctanh\beta} &
  &\implies&
  R_i &= \lambda R_i\partial_{\beta_i}\beta_i\arctanh\beta_i;
\end{align*}
that is, the optimum is achieved by setting all the $\beta_i$ equal to a constant, $\beta$. This gives $\dot H = 2R\beta\arctanh\beta=\frac{2R_c^2}{R} + \bigOO{\frac{R_c^4}{R^3}}$ and hence
\begin{align*}
  R_c \le \sqrt{\frac{PR}{2kT}}\,,
\end{align*}
leading once again to the scaling limit $R_c\lesssim\sqrt{AV}\sim V^{5/6}$, as the power $P$ scales with area and the gross computation rate $R$ with volume. Here we have presented a constructive proof of this scaling limit, as any valid reversible CRN implementation will yield this scaling limit.

\section{Relativistic Effects at Scale}
\label{sec:gr}

At different scales, from the microscopic to the cosmic, different constraints predominate in the analysis of maximum computation rate. For macroscopic systems at worldly scales, the aforementioned thermodynamic constraints are most directly relevant, yielding an upper bound of $R_C\lesssim\sqrt{AV}$. At sufficiently small scales, however, the surface area to volume ratio will be great enough that this thermodynamic constraint will no longer be limiting, with the volumetric Margolus-Levitin bound instead dominating. This reflects the fact that there is simply insufficient energy enclosed in the system for the resultant rate of computation to saturate the heat dissipation capacity at the boundary. If one were to use a denser computational architecture, the computational capacity of the region could be increased to the point that the power-flux bound is saturated, and thus recovering the more restrictive $\sqrt{AV}$ bound.

It transpires that there are two further regimes. As our systems approach cosmic scales, the threat of gravitational collapse becomes pertinent. In order to avoid this fate, we must lower the average density such that the system's radius always exceeds its Schwarzschild radius\footnote{In reality, the threshold radius for gravitational collapse exceeds the Schwarzschild radius by a factor not less than $\tfrac98$. The reasons for this shall be discussed in due course.}. As the Schwarzschild radius is proportional to the mass of the system, this implies that the computational rate of such large systems varies \textit{linearly} in radius. In fact, there is an intermediate regime: as we have not been utilising the full computational potential of our mass due to the thermodynamic constraint, we can gradually increase said utilisation whilst simultaneously reducing our overall density until the Margolus-Levitin limit is attained, at which point we default to the linear regime. This intermediate regime is thus still described by the $R_C\lesssim\sqrt{PM}$ limit, which in this post-Schwarzschild realm equates to $R_C\lesssim V^{1/2}$. In summary, the scaling laws of these four regimes go as $V$, $V^{5/6}$, $V^{1/2}$ and $V^{1/3}$, in order of increasing scale.

A more detailed analysis reveals that these large systems are subject to a further consideration. So far we have assumed a Galilean invariance worldview. At small and medium scales, this approximation is very good. At larger scales, however, relativistic effects threaten to reduce our overall computation rate via time dilation, and at even larger scales they can constrain the amount of mass we can fit within a volume lest the system undergo gravitational collapse as mentioned earlier. Special relativistic time dilation is easily avoided by minimising relative motion within the computational parts of the system (this implies that message packets will have reduced computational capacity compared to the static computational background, but message packets can generally be assumed static anyway).

Gravitational time dilation is a more serious issue, which cannot be eluded at large scales. In the case of a hypothetical supermassive computational system, local computation proceeds unabated, but distant users interfacing with the system will observe slower than expected operation, as depicted in \Cref{fig:constraint-gr}. To calculate this slowdown factor, we shall proceed by modelling the system as spherically symmetric. This is not unreasonable at these scales, where a body's self-gravitation makes maintaining other geometries unstable and impractical. We will not consider rotational systems which could allow for an ellipsoidal shape, as the requisite angular velocities would almost certainly abrogate nearly all computational progress solely due to special relativistic effects.

The metric\footnote{We use a $(+---)$ signature.} of an isotropic and spherically symmetric body in hydrostatic equilibrium can be obtained via the \emph{Tolman-Oppenheimer-Volkoff} (TOV) equation~\cite{tov-eqn},
\begin{equation}\label{eqn:tov-canonical}
\begin{split}
  \dv{P}{r} &= -\frac{Gm}{r^2}\rho\qty\Big(1+\frac{P}{\rho c^2})\qty\Big(1+\frac{4\pi r^3 P}{mc^2})\qty\Big(1-\frac{2Gm}{rc^2})^{-1} \\
  \dv{\nu}{r} &= -\qty\Big(\frac{2}{P+\rho c^2})\dv{P}{r} \\
      &= \frac{2Gm}{r^2c^2}\qty\Big(1+\frac{4\pi r^3P}{mc^2})\qty\Big(1-\frac{2Gm}{rc^2})^{-1} \\
  \dd s^2 &= e^\nu c^2 \dd t^2 - \qty\Big(1-\frac{2Gm}{rc^2})^{-1}\dd r^2 - r^2 \dd\Omega^2\,,
\end{split}
\end{equation}
where $P$ is the pressure, $m(r)$ is the mass enclosed by the concentric spherical shell of radius $r$, and $\rho$ is the local mass-energy density. All masses are as observed by a distant observer.

The rate of dynamics of a point $P$ as observed from a point $Q$ is well known to be
\begin{align*}
  \frac{R(Q)}{R(P)} &= \sqrt{\frac{g_{00}(P)}{g_{00}(Q)}}\,,
\end{align*}
where $g_{\mu\nu}$ is the metric tensor. We assume our distant observer to be moving slowly (relative to the computer) in approximately flat space, and thus that $g_{00}(Q)=1$. This yields a slowdown factor of $f(P)=\sqrt{g_{00}(P)}$. The total slowdown for an extended body is hence found to be
\begin{align*}
  f &= \frac1M \int_V \dd V \rho \sqrt{g_{00}} \\
    &= \frac1M \int_0^{r_1} \dd r \dv{m}{r} e^{\nu/2}\,, & \text{(spherically symmetric)}
\end{align*}
where $M$ is the total mass and $r_1$ is the least upper bound of its radial extent. 

In order to maximise $f$, we must clearly maximise $\nu$ throughout the body. It can be seen that $\dv{\nu}{r}\ge 0$, and in fact within solid regions where $\rho>0$ this inequality is strict. Furthermore, in empty space the TOV equation breaks down; instead, we use our Schwarzschild matching conditions to find
\begin{align*}
  \dv{\nu}{r} &= \frac{2Gm}{r^2c^2}\qty\Big(1-\frac{2Gm}{rc^2})^{-1} \\ \Delta\nu &= \Delta\log\qty\Big(1-\frac{2Gm}{rc^2})\,,
\end{align*}
which shows that this inequality is always strict whenever $m>0$. $\nu$ at the surface is fixed by the Schwarzschild metric, in accordance with Birkhoff's theorem, and thus for a fixed extent our only control over $\nu$ is the distribution $m$ between the surface and core. Approaching the core from the surface, $\nu$ strictly decreases until $m$ vanishes. Furthermore $\dd\nu/\dd r$ is at least as great as in empty space, being strictly greater in massive regions. Therefore it is desirable to concentrate mass towards the surface.

We assume that the density of our architecture of choice is bounded from above, thus limiting the degree to which we can concentrate mass towards the surface. The optimum is then achieved by a thick shell of limiting density from the surface inwards. The two extreme cases of a thin shell and a solid sphere can be treated exactly, but an arbitrarily thick spherical shell must be treated numerically.

\para{Solid Sphere}

A solid sphere coincides with the Schwarzschild geometry. The usual Schwarzschild metric corresponds to the exterior of the object, instead we must use the interior solution which can be derived from the TOV equation~(\ref{eqn:tov-canonical}), yielding
\begin{align*}
  \sqrt{g_{00}} &= \frac32\sqrt{1-\frac{r_s}{r_1}} - \frac12\sqrt{1-\qty\Big(\frac{r}{r_1})^2\frac{r_s}{r_1}}
\end{align*}
where $r_s=2GM/c^2$ is the Schwarzschild radius. Before computing $f$, note that $g_{00}$ vanishes when $r_s=\tfrac89r_1$. This limit is in fact of great importance; whilst an object of size $r_1=\tfrac98 r_s$ would exceed the Schwarzschild radius and be presumed stable against gravitational collapse, this is not the case. At this point, such an object is unstable towards spontaneous oscillations, its core pressure diverges, and it readily collapses to a black hole~\cite{schwarz-lim-main}. Furthermore, depending on its heat capacity ratio $\gamma$, a gaseous hydrostatic sphere may be unstable at even larger radii as tabulated by \textcite{schwarz-lims}. This $r_1\ge\tfrac98r_s$ threshold is thus a stronger bound on the size of massive objects. Integrating $\sqrt{g_{00}}$ over the mass of this solid sphere, we find
\begin{align*}
  f_{\text{solid}} &= \frac{3}{16 v_s}( (1+6v_s)\sqrt{1-v_s} - v_s^{-1/2}\arcsin\sqrt{v_s} )
\end{align*}
where we have introduced the normalised coordinate $v=r/r_1$. At the $v_s=\tfrac89$ threshold, we get $f_{\text{solid}}\approx 0.1699$.

\para{Thin Shell}

For a thin shell, we first identify that the pressure on the outer boundary vanishes. Let the inner radius be $r_0$ and the thickness be $\delta r=r_1-r_0$ such that $M\approx 4\pi\rho r_1^2\delta r$ and $\delta v\ll 1$. We also introduce the unitless variable $u=P/\rho c^2$,
\begin{align*}
  \dv{u}{r} &= -\frac{4\pi\rho rG}{c^2}(1+u)\qty\Big(\frac{r-r_0}{r} + u)\qty\Big(1-\frac{2Gm}{rc^2})^{-1} \\
  \dv{u}{v} &= -\frac{4\pi\rho rr_1 G}{c^2}(1+u)\qty\Big(\frac{r-r_0}{r} + u)\qty\Big(1-\frac{2GM}{c^2}\frac{r-r_0}{r\delta r})^{-1} \\
    &\approx -\frac{v_s}{2\delta v}(u+1)(u+\beta\delta v)(1-\beta v_s)^{-1} \\
  \Delta u &\approx -\tfrac12v_s(u+1)(u+\beta\delta v)(1-\beta v_s)^{-1}\,,
\end{align*}
where $\beta = \frac{r-r_0}{r_1-r_0} \in[0,1]$. Integrating from the surface inwards, we have $u_1=0$ and $\beta=1$, yielding
\begin{align*}
  u_0 &= \frac{\delta v}{2} \frac{v_s}{1-v_s} + \bigO{\delta v^2}\,.
\end{align*}
Therefore in the limit $\delta v\to 0$, the pressure throughout the thin shell vanishes. As a result, we also have that $\nu$ is constant for all $v\in[0,1]$, taking the value $1-v_s$ and giving
\begin{align*}
  f_{\text{thin}} &= \sqrt{1-v_s}\,.
\end{align*}
At the threshold, we get $f_{\text{thin}}=\tfrac13$.

\para{Thick Shell}

For general $\delta v$, we integrate numerically. We rewrite the TOV equations in normalised and unitless form thus
\begin{align*}
  \dv{\log g}{v} &= \frac{vv_s}{2}\frac{\mu+3u}{\mu_1-v^2v_s\mu}\,, &
  \dv{u}{v} &= -(1+u)\dv{\log g}{v}\,, &
  \dv{f}{v} &= -\frac{3v^2g}{\mu_1}\,,
\end{align*}
where $\mu=1-(v_0/v)^3$, $\mu_1=1-v_0^3$ and $g\equiv \sqrt{g_{00}}$. We integrate the system from $v=1$ down to $v=v_0$, with initial conditions $u_1=0$, $g_1=\sqrt{1-v_s}$ and $f_1=0$. Our slowdown factor is given by $f_0$, and explicit results are given in the following section.

\section{Variance in Performance at Different Scales}

In order to calculate the maximum performance of a system at different sizes, we compute the optimum speed within each constraint---thermodynamic and Margolus-Levitin---with relativistic corrections, and then pick the least of these upper bounds. To apply the relativistic constraint, we substitute $M=\tfrac12 r_s c^2 / G$ for mass where $r_s$ is the Schwarzschild radius, and then multiply by our slowdown factor $f$. For better parametericity, we actually use $r_s=v_s \ell$ where $\ell$ is the system's radius/linear dimension. This yields the two bounds,
\begin{align*}
  R_{\text{thermo.}} &= \alpha \sqrt{\ell^3 v_s} f(v_s,v_0)\,, &
  R_{\text{marg.lev.}} &= \beta \ell v_s f(v_s,v_0)\,,
\end{align*}
where $\alpha$ and $\beta$ are constants of proportionality that depend on the system architecture. Notice that there is in fact some choice available in system geometry; we can pick any constant density thick shell solution, with parameters $(v_0,v_s)$. As we increase $v_s$, the factor $f$ will decrease, and so there will be an optimum pair $(v_0,v_s)$ maximising the given bound. The maximum density constraint is equivalent to setting a maximum value to the mass and hence $v_s$,
\begin{align*}
  v_s &= \frac1\ell \frac{2GM}{c^2} = \ell^2\frac{8\pi\rho G}{3c^2}(1-v_0^3) \\
    &\le v_m \equiv \ell^2\frac{8\pi\rho G}{3c^2} \\
    &\le \frac89\,,
\end{align*}
where the last constraint is to avoid gravitational collapse. For a given $v_m$ then, we have $v_0=\sqrt[3]{1-v_s/v_m}$ and we maximise the two $R$ bounds in $0<v_s\le\min(\tfrac89,v_m)$.

An illustrative example of these regimes is depicted in \Cref{fig:lims}. Notice how the solid sphere, though maximising usable computational matter and thus its local computational rate, has its observable computational rate abrogated as it approaches the point of gravitational self-collapse. This demonstrates how reducing the available matter can enable the observable computational rate to increase, though the difference between the dashed and solid lines shows that there remains a relativistic penalty in the form of gravitational time dilation. We also see that an irreversible system underperforms with respect to a reversible system across a large range of length scales, otherwise matching it at the extremes. The exact range again depends on system architecture specifics, and for some architectures the irreversible system may give transient superior performance at certain scales.

{
\newlength{\limexsize}
\setlength{\limexsize}{4.2cm}
\newcommand*{\limex}[1]{
  \begin{subfigure}[t]{\limexsize}
    \centering
    \caption{}\label{fig:lims-#1}
    \input{fig-lims-sub-#1}
  \end{subfigure}}
\newcommand*{\limkey}{\parbox[t]{3\limexsize}{\hfill
\begingroup
  \makeatletter
  \providecommand\color[2][]{%
    \GenericError{(gnuplot) \space\space\space\@spaces}{%
      Package color not loaded in conjunction with
      terminal option `colourtext'%
    }{See the gnuplot documentation for explanation.%
    }{Either use 'blacktext' in gnuplot or load the package
      color.sty in LaTeX.}%
    \renewcommand\color[2][]{}%
  }%
  \providecommand\includegraphics[2][]{%
    \GenericError{(gnuplot) \space\space\space\@spaces}{%
      Package graphicx or graphics not loaded%
    }{See the gnuplot documentation for explanation.%
    }{The gnuplot epslatex terminal needs graphicx.sty or graphics.sty.}%
    \renewcommand\includegraphics[2][]{}%
  }%
  \providecommand\rotatebox[2]{#2}%
  \@ifundefined{ifGPcolor}{%
    \newif\ifGPcolor
    \GPcolortrue
  }{}%
  \@ifundefined{ifGPblacktext}{%
    \newif\ifGPblacktext
    \GPblacktexttrue
  }{}%
  \let\gplgaddtomacro\g@addto@macro
  \gdef\gplbacktext{}%
  \gdef\gplfronttext{}%
  \makeatother
  \ifGPblacktext
    \def\colorrgb#1{}%
    \def\colorgray#1{}%
  \else
    \ifGPcolor
      \def\colorrgb#1{\color[rgb]{#1}}%
      \def\colorgray#1{\color[gray]{#1}}%
      \expandafter\def\csname LTw\endcsname{\color{white}}%
      \expandafter\def\csname LTb\endcsname{\color{black}}%
      \expandafter\def\csname LTa\endcsname{\color{black}}%
      \expandafter\def\csname LT0\endcsname{\color[rgb]{1,0,0}}%
      \expandafter\def\csname LT1\endcsname{\color[rgb]{0,1,0}}%
      \expandafter\def\csname LT2\endcsname{\color[rgb]{0,0,1}}%
      \expandafter\def\csname LT3\endcsname{\color[rgb]{1,0,1}}%
      \expandafter\def\csname LT4\endcsname{\color[rgb]{0,1,1}}%
      \expandafter\def\csname LT5\endcsname{\color[rgb]{1,1,0}}%
      \expandafter\def\csname LT6\endcsname{\color[rgb]{0,0,0}}%
      \expandafter\def\csname LT7\endcsname{\color[rgb]{1,0.3,0}}%
      \expandafter\def\csname LT8\endcsname{\color[rgb]{0.5,0.5,0.5}}%
    \else
      \def\colorrgb#1{\color{black}}%
      \def\colorgray#1{\color[gray]{#1}}%
      \expandafter\def\csname LTw\endcsname{\color{white}}%
      \expandafter\def\csname LTb\endcsname{\color{black}}%
      \expandafter\def\csname LTa\endcsname{\color{black}}%
      \expandafter\def\csname LT0\endcsname{\color{black}}%
      \expandafter\def\csname LT1\endcsname{\color{black}}%
      \expandafter\def\csname LT2\endcsname{\color{black}}%
      \expandafter\def\csname LT3\endcsname{\color{black}}%
      \expandafter\def\csname LT4\endcsname{\color{black}}%
      \expandafter\def\csname LT5\endcsname{\color{black}}%
      \expandafter\def\csname LT6\endcsname{\color{black}}%
      \expandafter\def\csname LT7\endcsname{\color{black}}%
      \expandafter\def\csname LT8\endcsname{\color{black}}%
    \fi
  \fi
    \setlength{\unitlength}{0.0500bp}%
    \ifx\gptboxheight\undefined%
      \newlength{\gptboxheight}%
      \newlength{\gptboxwidth}%
      \newsavebox{\gptboxtext}%
    \fi%
    \setlength{\fboxrule}{0.5pt}%
    \setlength{\fboxsep}{1pt}%
\begin{picture}(5660.00,1120.00)%
    \gplgaddtomacro\gplbacktext{%
    }%
    \gplgaddtomacro\gplfronttext{%
      \csname LTb\endcsname
      \put(4915,650){\makebox(0,0)[r]{\strut{}Quantum~~~}}%
      \csname LTb\endcsname
      \put(4915,475){\makebox(0,0)[r]{\strut{}Thermo (Irrev.)~~~}}%
      \csname LTb\endcsname
      \put(4915,300){\makebox(0,0)[r]{\strut{}Thermo (Rev.)~~~}}%
      \csname LTb\endcsname
      \put(4915,125){\makebox(0,0)[r]{\strut{}Grav. Threshold.~~~}}%
    }%
    \gplbacktext
    \put(0,0){\includegraphics[width={283.00bp},height={56.00bp}]{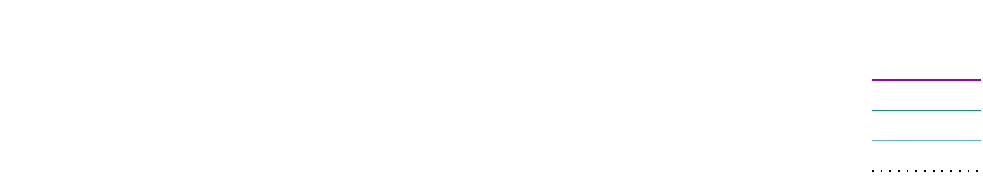}}%
    \gplfronttext
  \end{picture}%
\endgroup
}}
\newcommand*{\limsp}{\phantom{\parbox[t]{\limexsize}{~}}}
\begin{figure}[p]
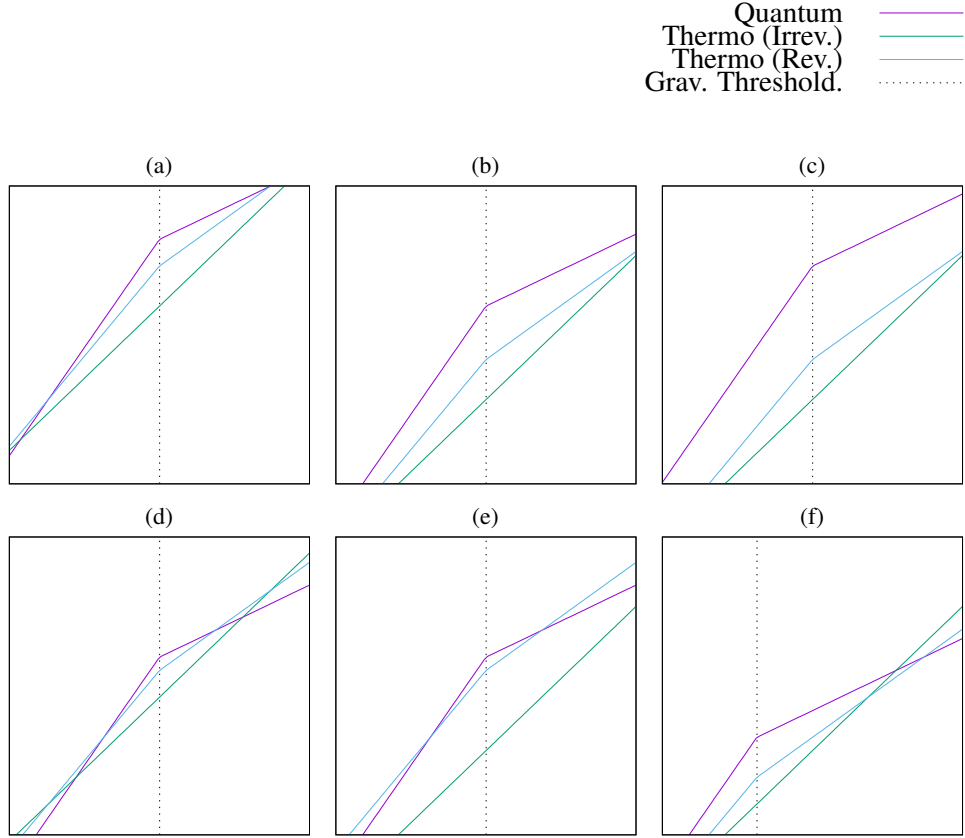

  \centering
  \parbox[t]{3\limexsize}{\hfill\input{fig-lims-sub-key}}
  
  \begin{subfigure}[t]{\limexsize}
    \centering
    \caption{}\label{fig:lims-1}
    \input{fig-lims-sub-1}
  \end{subfigure}
  \begin{subfigure}[t]{\limexsize}
    \centering
    \caption{}\label{fig:lims-2}
    \input{fig-lims-sub-2}
  \end{subfigure}
  \begin{subfigure}[t]{\limexsize}
    \centering
    \caption{}\label{fig:lims-3}
    \input{fig-lims-sub-3}
  \end{subfigure}
  
  \begin{subfigure}[t]{\limexsize}
    \centering
    \caption{}\label{fig:lims-4}
    \input{fig-lims-sub-4}
  \end{subfigure}
  \begin{subfigure}[t]{\limexsize}
    \centering
    \caption{}\label{fig:lims-5}
    \input{fig-lims-sub-5}
  \end{subfigure}
  \begin{subfigure}[t]{\limexsize}
    \centering
    \caption{}\label{fig:lims-6}
    \input{fig-lims-sub-6}
  \end{subfigure}
  \caption{In these plots are shown a series of idealised examples of the relevant bounds on computation rate. The vertical axes represents rate of computation in arbitrary logarithmic units, and the horizontal axes the radius of the system in arbitrary logarithmic units. The axis bounds are the same for all Figures (a--g). The examples are parameterised by mass-energy density, raw rate of computation per unit mass, power flux through the convex bounding surface, entropy cost of erasing one bit of information, and temperature of the system.\capnl%
  For a given density, there will exist a threshold radius at which the Schwarzschild radius and system radius of a solid sphere coincide, leading to gravitational collapse. This threshold is indicated by the vertical dashed line. As discussed, the geometry must transition to a spherical shell beyond this point. The lines labelled `quantum' correspond to the quantum mechanical upper bound on computation rate, which scales as $M\sim R^3$ below the threshold radius, and $M\sim R$ beyond. The lines labelled `thermo (irrev.)' correspond to the thermodynamic upper bound on the rate of irreversible computation, which scales as $A\sim R^2$. The lines labelled `thermo (rev.)' correspond to the thermodynamic upper bound on the rate of reversible computation, scaling as $R^{2.5}$ and $R^{1.5}$ below and above the threshold radius respectively. (continued)}\label{fig:lims}
\end{figure}
\begin{figure}[p]\ContinuedFloat
  \centering
  \begin{subfigure}[t]{\linewidth}
    \centering
    \caption{}\label{fig:lims-main}
    \input{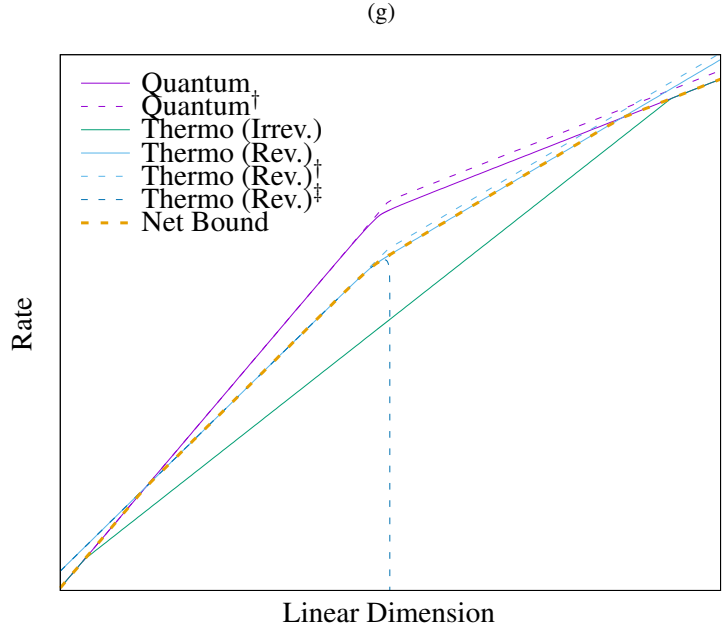}
  \end{subfigure}
  \caption{(continued) In each of these examples, assume the same parameter values unless otherwise specified: (a)~the maximum possible performance, as dictated by the Margolus-Levitin and Szilard-Landauer bounds; (b)~the rates for the particular architecture as characterised by the parameter values; (c)~higher raw rate of computation per unit mass (bounded by $c^2/h_P$); (d)~greater power flux or lower entropy cost per erased bit (bounded by $k_B\log 2$); (e)~lower temperature; (f)~higher density.\capnl%
  For each of these examples, the maximum computation rate for a given system size is given by the least of the three upper bounds. These figures are also idealised in the sense of ignoring gravitational time dilation for simplicity.\capnl%
  A more complete picture is given in (g). The solid lines show constraints corrected for time dilation$^\ast$, whilst the dashed variants ($\dagger$) are uncorrected variants and the $(\ddagger$) variant shows the case without geometric optimisation (i.e.\ maintaining a solid sphere of constant density rather than transitioning to a spherical shell). The dashed `net bound' line shows the best rate achievable for a given system radius.}
  ~\\\parbox{\textwidth}{\hangindent=0.3cm $^\ast$\scriptsize Time dilation corrections were computed by integrating the thick shell system of differential equations using \texttt{dopri5}, and the shell thickness was optimised to maximise the rate using a golden section search. The code was written in \texttt{haskell} and is available, complete with integration and optimisation routines, at \url{https://github.com/hannah-earley/revcomp-relativistic-limits}.}
\end{figure}
}

\section{Discussion}

\para{Additional Entropic Sources}

The preceding analysis has been restricted to dissipating entropy arising in the course of the computation itself. However, a physical computer will be subject to other sources of entropy in addition to this. As mentioned in the introduction, Bennett calculated that the influence of turbulence in the atmospheres of nearby stars would be sufficient to thermalise a billiard ball computer within a few hundred collisions. Thus, in general, we must aim to shield our computer from such external influences, for example by error correction procedures. Fortunately, the influences of such external sources may only interact with our system via the same boundary through which we dissipate computational entropy, and as such their rate is at most proportional to this surface area\footnote{Even though such external influences may indeed permeate the entire volume, such as gravitational waves or neutrinos, the \emph{rate} of incidence of these interaction will necessarily scale areametrically. Moreover, gravitational waves and neutrinos only penetrate the entire volume because they are so weakly interacting. A more strongly coupled interaction such as ultraviolet light or particulate matter will primarily affect only the `skin' of the system, and thus illustrates how such external influences are genuinely only areametric in strength.}. Thus, as long as the external sources are not overwhelmingly strong we should be able to suppress them at all scales with equal ease.

A more challenging source of errors and entropy comes from within. Assuming a non-vanishing system temperature, the entire system volume will generate thermodynamic fluctuations at a commensurate rate. These fluctuations can manifest in a variety of ways, perhaps the most damaging of which is radioactive decay leading to radical generation with the potential to damage the computational structure. Every event leading to a change in computational state or requiring repair invokes an entropic cost, and given that these events occur with a rate proportional to the computationally active volume, this ultimately recovers an areametric bound to our computational rate,
\begin{align*}
  V &\le \frac{P}{kT}\frac1{\dot\eta_{\text{int.fluc.}}}\,.
\end{align*}

This does not necessarily render our reversible computation performance gains unattainable, however. Providing $\dot\eta_{\text{int.fluc.}}$ is sufficiently small, we can outperform irreversible computers at scales up to
\begin{align*}
  \ell &\le \sqrt{\frac{P}{kT}\frac1{\dot\eta_{\text{int.fluc.}}}\frac1{\delta r}}
\end{align*}
where $\delta r$ is the thickness of the irreversible computational shell that can be supported by such an architecture. If this threshold size is sufficiently large as to coincide with the Schwarzschild threshold of \Cref{fig:lims} then such a reversible architecture would in fact outperform its irreversible analogue at all scales. Furthermore, it is likely that $\dot\eta_{\text{int.fluc.}}$ is smaller than it would be in an irreversible system as they have one fewer source of fluctuations to combat; namely, irreversible computers expend significant quantities of energy to ensure unidirectional operation, and any deviation from this is a potentially fatal error. In contrast, a reversible computer is intrinsically robust to such `errors', particularly the Brownian architecture described, which actively exploits this. Expending less energy has the added advantage of permitting lower operating temperatures, further reducing $\dot\eta_{\text{int.fluc.}}$.

\para{Mixed Architectures}

Because of the generality of the $R\lesssim\sqrt{AV}$ scaling law, systems of maximal computational rate can be constructed using any desired mix of architectures in order to meet the requirements of a given problem. Providing entropy and input power can be efficiently transported between the surface and the computational bulk, the entropy generation and power constraints superpose linearly. Thus in principle it is possible to have a mix of quantum and Brownian architectures within a single system. A caveat is that these two architectures may need to operate at different temperatures, and a temperature gradient introduces an additional source of entropy. Thus, the area of the boundary between these subsystems must be at most proportional to the system's bounding area in order that the entropy generated can be effectively countered.

This principle is more general: any non-equilibrium inhomogeneity in the system's structure will result in entropy generation. To quantify this, we assume that such inhomogeneities equilibrate via an uncorrelated diffusive process, and we proceed via a Fokker-Planck approach.

\let\grad\nabla
The Fokker-Planck equation describes the evolution of a probability distribution in space due to stochastic dynamics. Specifically, it pertains to the influence of a Langevin force with rapidly decaying correlations in time. In this limit, Brownian particles will exhibit a combination of drift and diffusion depending on the properties of the system. The evolution of the probability distribution $\varphi$ is found~\cite{risken} to obey\footnote{In \textcite{risken}, the convention is $\dot\varphi=-\partial_i\mu_i\varphi+\partial_i\partial_jD_{ij}\varphi$. We use a different convention here for convenience, wherein $\mu_i'=\mu_i-\partial_jD_{ij}$.}
\begin{align*}
  \dot\varphi &= -\grad\cdot [ \mu\varphi - D\grad\varphi ]
\end{align*}
where $\mu$ is a drift vector and $D$ is a diffusion matrix. We also identify the probability current $\mu\varphi-D\grad\varphi$. In order to establish the rate of entropy generation, we must first determine the steady state distribution. We make the assumption that the steady state probability current is everywhere zero, i.e.\ there are no persistent current flows or vortices, and therefore find that
\begin{align*}
  \frac{\nabla\varphi_0}{\varphi_0} &= D^{-1}\mu\,.
\end{align*}

Now we obtain an expression for the entropy of the system. As there are many particles, we can reinterpret $\varphi$ as a concentration of particles (as the Fokker-Planck equation does not impose any normalisation), and so each particle will have an associated entropy of $1+\varepsilon-\log\lambda$ where $\lambda\propto\varphi$. This yields intensional entropy $\eta=\varphi(1+\varepsilon-\log\lambda)$ and $\dot\eta=\dot\varphi(\varepsilon-\log\lambda)$. At equilibrium, $\dot\eta$ cannot have a leading order linear term in $\varphi$, and so $\varepsilon=\log\lambda_0$ which gives
\begin{align*}
  \eta &= \varphi\qty\Big(1-\log\frac{\varphi}{\varphi_0}) \\
  H &= \int \dd V\, \varphi\qty\Big(1-\log\frac{\varphi}{\varphi_0}) \\
  \dot H &= -\int \dd V\, \dot\varphi\log\frac{\varphi}{\varphi_0}\,.
\end{align*}

We now substitute the Fokker-Planck equation for $\dot\varphi$, obtaining
\begin{align*}
  \dot H &= \int \dd S \cdot (\mu\varphi - D\grad\varphi) \log\frac{\varphi}{\varphi_0} - \int \dd V\, (\mu\varphi - D\grad\varphi) \cdot\grad\log\frac{\varphi}{\varphi_0}  \\
    &= \int\dd V\, (D\grad\varphi-\mu\varphi)\cdot\grad\log\frac{\varphi}{\varphi_0} \\
    &= \int\dd V\, D\varphi\qty\Big(\frac{\grad\varphi}\varphi - D^{-1}\mu)\cdot\grad\log\frac{\varphi}{\varphi_0} \\
    &= \int\dd V\, D\varphi\qty\Big(\frac{\grad\varphi}\varphi - \frac{\grad\varphi_0}{\varphi_0})\cdot\grad\log\frac{\varphi}{\varphi_0} \\
    &= \int\dd V\, \varphi\qty\Big(D\grad\log\frac{\varphi}{\varphi_0})\cdot\grad\log\frac{\varphi}{\varphi_0} \\
  \dot H &= \sum_i N_i \evqty\Big{\,\qty\Big| D_i^{1/2} \grad\log\frac{\varphi_i}{\varphi_{i,0}}|^2}\,,
\end{align*}
where $N_i$ is the number of particles of diffusive species $i$. In the second line we have assumed that the probability current across the system boundary vanishes and in the last line we have generalised to multiple diffusive species.

The consequence is that a system may only sustain the use of spatial variation in its architecture if at least one of the following three conditions holds for each such species,
\begin{enumerate}
  \item The region of variation scales with the system's surface area, such as if the gradient is localised to a hemispheric plane,
  \item The diffusion rate $D$ is vanishingly small,
  \item The average relative strength of the gradient, $\grad\log(\varphi/\varphi_0)$, has scaling no greater than $\ell^{-1/2}\sim V^{-1/6}$.
\end{enumerate}
These conditions may be violated only to the extent that another condition is exploited to achieve a commensurate reduction in the entropy rate.

\para{Modes of Computation}

The quantum Zeno architecture is inherently processive, meaning that every transition it makes is useful. This is in contrast to the Brownian architecture, in which only a vanishingly small fraction of transitions lead to a net advance in computational state. There are a number of consequences to this distinction that make the quantum architecture preferable.

Firstly, a quantum architecture equivalent to a given Brownian architecture will only use $N_{\text{QZE.}}\sim\sqrt{AV}$ active computational elements compared to $N_{\text{Brown.}}\sim V$, which significantly reduces the risk of errors due to internal fluctuation. In addition, the properties of the QZE mean that the fluctuations within the computational subvolume of the quantum architecture can be rectified at all scales in contrast to the classical system. Of course, fluctuations in the remaining bulk of the volume are still problematic, but their rectification is less critical.

Secondly, whilst the two systems would have the same net rate of computation, the computational elements of the quantum system operate at the same maximum speed regardless of scale, and the degree of parallelism can be tuned between maximally parallel ($N\sim\sqrt{AV}$) and fully serial ($N=1$). This means that the quantum architecture can operate equally well for parallel and serial problems. One caveat is that the maximal attainable parallelism of the quantum system, $\sim\sqrt{AV}$, is lower than for the Brownian, $\sim V$. In practice this is not too problematic as parallel problems can be simulated serially, and because the total computation rate is independent of the degree of parallelism, the quantum system experiences no penalty for this choice. For the Brownian system, however, even a problem maximally exploiting the available parallelism will be limited by the time for each element to perform a single net operation. In addition, whilst the quantum system has fewer computational foci than the Brownian system, the remaining bulk can be used for `cold' computation such as data storage.

Thirdly, almost all parallel problems will involve inter-element communication and synchronisation. As will be discussed in the next Part~\cite{earley-parsimony-ii}, these processes present significant challenges in the Brownian architecture that in the worst case would throttle the system down to $R\sim A$. This arises because synchronisation processes map to constrictions in phase space, and Brownian diffusion through such a constriction is very slow. As the quantum architecture evolves processively, it is less subject to this effect and thus its computational elements can communicate much more freely.

Fourthly, the quantum architecture permits quantum computation whereas the Brownian architecture is unlikely to be capable of supporting quantum computation. The reason for this is that the average time for each computational transition in the Brownian system is typically large, increasing as the system size does, whereas the decoherence timescale for a quantum state is fixed and typically small. Furthermore, the decoherence timescale is likely significantly smaller than for the QZE architecture as the Brownian system probably operates at a higher temperature. Altogether, these factors make it incredibly unlikely that a quantum state can be reliably maintained through a quantum computation within such an environment.

Despite all these disadvantages, the Brownian architecture is perhaps of more interest as it is ostensibly easier and cheaper to construct with current technology. The most exciting avenue for this may lie in the fields of biological and molecular computing, in which molecules such as DNA are constructed in such a way that their resulting dynamics encode a computational process~\cites{winfree-tam,cardelli-dsd}. Chemical systems are a natural substrate for Brownian dynamics, and the manipulation of DNA is becoming ever cheaper and more sophisticated. With improvements in synthetic biology, we may even be able to readily prepare self-repairing DNA computers in the near future.

We finish by discussing the timestep for a net computational transition in the Brownian architecture. This timestep is controlled by the biases, $b_i$. To maximise the system's computation rate, we take $b_i=\beta\propto1/\sqrt\ell$. If, at operating reactant concentrations, the time for any computational step (forwards or backwards) is $t_0$, then the net timestep is given by 
\begin{align*}
  \tau &= \frac{t_0}{\beta} \propto \ell^{1/2} \sim V^{1/6}
\end{align*}
which can be seen to get larger as the system gets larger. Fortunately this scaling is sublinear, and so depending on the available power, the achievable bias may be significant. For example, assuming a \SI{500}{\watt\per\meter\squared} radiative capacity and a \SI{1}{\meter} radius system consisting of fairly conservative computational particles of size $(\SI{10}{\nano\meter})^3$ operating at a gross rate of \SI{1}{\hertz}, a bias as high as $\beta\sim0.4$ is possible.

Another approach to managing modes of computation in a Brownian computer is to institute a hierarchy of bias levels at different concentrations. For example, a viable system may consist of the following tuples $\sim(N,b;R_C)$,
\begin{align*}
  \{ (V,\ell^{-1/2};V^{5/6}),~ (V^{14/15},\ell^{-2/5};V^{4/5}),~ (V^{5/6},\ell^{-1/4};V^{3/4}),~ (V^{2/3},1;V^{2/3}) \}
\end{align*}
and computations requiring faster steps may use higher biases at the expense of less computational capacity. Additionally, the highest bias level admits irreversible computation, and so we may enclose our computer with a thin irreversible shell, whilst the internal bulk consists of reversible computations at various bias levels. Going further, we could have a hot inner Brownian core, a cold outer quantum core, and an enveloping irreversible silicon shell. Additionally, a hierarchical Brownian system need not employ spatial variation; by using different species for each bias, the subsystems can be mixed homogeneously.

\section{Conclusion}

We have shown that, subject to reasonable geometric constraints on power delivery and heat dissipation, the rate of computation of a given convex region is subject to a universal bound, scaling as $R_C\lesssim\sqrt{AV}\sim V^{5/6}$, and that this can only be achieved with reversible computation. For irreversible computation, this bound falls to $A\sim V^{2/3}$. The scaling laws persist at all practical scales, only breaking down when the system size approaches the Schwarzschild scale. For typical densities ($\sim\SI{1000}{\kilo\gram\per\meter\cubed}$), this threshold scale is $\SI{4e11}{\meter}\approx\SI{2.7}{AU}$. Beyond this scale, the maximum rate is attained by localising mass into as thin a shell as possible, reminiscent of megastructures such as Dyson spheres from the world of science fiction. At very small scales, the rate scales with $V$ as the surface area to volume ratio is negligible. The scaling then falls to $V^{5/6}$, to $V^{1/2}$, and finally to $V^{1/3}$ as the system increases in size. At the Schwarzschild regime and beyond, the computation rate is suppressed by a factor of order unity due to gravitational time dilation.

This analysis has assumed that each computational element acts independently. In our next two papers we shall investigate the constraints affecting cooperative reversible architectures, in particular the thermodynamic cost of synchronisation processes, such as communication and resource sharing. These costs turn out to be quite significant, even prohibitive.

\appendix

\section{Acknowledgements}

The author would like to acknowledge the invaluable help and support of his supervisor, Gos Micklem. This work was supported by the Engineering and Physical Sciences Research Council, project reference 1781682.

\section{Reversible Computing}
\label{app:rev-comp}

Irreversible computers are characterised by their non-conservation of information, often in subtle ways. This may be considered to ultimately originate in early mathematical models of computations such as the Turing Machine and Lambda Calculus in which overwriting and discarding, respectively, are implicit to the primitive operations of the models. This implicitness is pervasive across practically all levels of abstraction. At the lowest levels, this manifests in such forms as transistor logic, non-injective logic gates such as \texttt{AND}, the free overwriting of memory locations and registers, and jumping to other instructions without a trace of the instruction pointer's origin.
Semiconductor transistors and logic gates are intrinsically non-invertible because the state(s)---conductive or not---are properties of the semiconductor regions themselves; therefore, each time the semiconductor state changes in response to its input, it must first forget its previous state.
At higher levels it is even less obvious, from variables going out of scope, to ignoring the return value of a function, to iteration and recursion without tracking the full history of such processes, to pure functional languages encouraging rapid turnover of memory through an underlying garbage collector.

With such ubiquitous application of information-destructive primitives, it is hard to see how algorithms can be rewritten reversibly. In an upcoming paper, we shall introduce a model of reversible computation appropriate for the Brownian architecture described earlier, as well as give an overview of other extant reversible programming languages such as \emph{Janus}~\cite{janus}, \emph{$\Psi$-Lisp}~\cite{psi-lisp} and \emph{Theseus}~\cite{theseus}. Our Brownian language is high level, demonstrating that reversible programming need not be overtly difficult. Adopting this paradigm requires a shift in thinking, in the same way as moving between imperative and functional styles of programming does. 

To program reversibly requires some care. In particular, it is not sufficient for it to be possible in principle to reconstruct the computational history; instead each computational primitive must be intrinsically invertible such that the computational history can be readily rewound step-by-step as easily as it was run forward. Consider a branch in control flow due to a conditional statement; it is essential that the information used to switch between the branches be retained, and furthermore that immediately following the branch this information is sufficiently accessible for us to immediately step backwards, reversing the conditional and merging the control flow. This prohibits, for example, branching on the result of a transient return value from a function; this value must be retained.

After branching the control flow, one often wishes to merge control flow again. It is imperative that after such an operation it is explicitly known which branch was taken, typically in the form of a variable value. In fact, the resolution to this problem showcases the elegant symmetry of this paradigm: the merging of control flow is simply the inverse of a branch, so one simply takes a condition that distinguishes the two branches, and plugs it in in reverse. Going further, we can introduce reversible iteration; a reversible loop has two branch points: an entry point where control flow either enters the loop or continues to the next iteration, and an exit point where control flow either continues or exits the loop. The conditions are then, respectively, whether or not this is the first iteration and whether or not this is the last iteration.

More concrete operations, such as an arithmetic primitive for addition, must also be rendered reversible. Clearly the operation $(x,y)\mapsto(x+y)$ is not invertible as, given $6$, it is unclear whether the addends were $(2,4)$ or $(1,5)$. In such cases, one often finds that there are multiple ways to render such irreversible primitives reversible. Here, two possible implementations are $(x,y)\mapsto (x,x+y)$ and $(x,y)\mapsto(x-y,x+y)$.

When rewriting a program in this way, one often finds that the `additional' information thus obtained can be recycled elsewhere, reducing the need for recalculation or copying of values, though such parsimonious algorithms are not always easy to identify. Fortunately if it is desired to discard a value, this does remain possible; one can reversibly discard it by using one of Bennett's algorithms~\cites{bennett-tm,bennett-pebbling}, or one can introduce an explicit discard primitive to send it to an entropy dump. Though the latter option may seem antithetical, its explication at least makes clear where sources of entropy arise in the program, and thus where opportunities for optimisation may exist. What Bennett showed, however, is that dumping entropy is never necessary. His simplest algorithm shows how an irreversible program $P$ can be embedded reversibly; suppose that we keep track of the data discarded by $P$, then when given an input $x$, $P$ maps it to the tuple $(y,h)$ where $y$ is the output of $Px$ and $h$ is the `history' data. Now, make a copy of $y$, $y'$, and set it aside. We can now run $P$ backwards as we have its discarded data, $P^{-1}:(y,h)\mapsto x$. Therefore we are left with $(x,y')$, i.e. a reversible embedding of $P$, $P':x\mapsto (x,Px)$. The discarded data $h$ may be quite large, but Bennett developed more sophisticated algorithms which limit the intermediate history data recorded to $\bigOO{\log t}$ where $t$ is the program runtime.

\printbibliography
\end{document}